\documentclass[11pt]{article}

\usepackage{amsmath}

\newcommand{\bfbeta}{\mbox{\boldmath $\beta$}}

		\newcommand{\bfx}{{\bf x}}	
		\newcommand{\bfz}{{\bf z}}

\RequirePackage{amsthm,amsmath,amsfonts,amssymb}
\RequirePackage[numbers,sort&compress]{natbib}
\RequirePackage[colorlinks,citecolor=blue,urlcolor=blue]{hyperref}
\RequirePackage{graphicx}
\usepackage{natbib}
\usepackage{enumitem}
\usepackage{outlines}
\usepackage{booktabs} %for \toprule \bottomrule in tables
\usepackage{array}

\usepackage{newtxtext,newtxmath}  % Times font

\usepackage{float}  
\usepackage{titlesec}  % Allows customization of section titles
\titleformat{\section}  % Customize section headings
  {\centering\bfseries\Large}  % Centered, bold, large font
  {\thesection.}  % Include section numbering (e.g., "1. INTRODUCTION")
  {0.5em}  % Space between number and title
  {\MakeUppercase}  % Convert title to uppercase
\theoremstyle{plain}

\theoremstyle{remark}

\newcommand{\tb}{\textbf}

\title{{\bfseries \Huge Comparative Review of Modern Competing \\[.1em]Risk Methods in High-dimensional Settings}}
\author{
  \Large \tb{Paul M. Djangang}\textsuperscript{1},  
  \tb{Summer S. Han}\textsuperscript{2},  
  \tb{Nilotpal Sanyal}\textsuperscript{1}\thanks{Corresponding author. Email: nsanyal@utep.edu} 
}

\date{}
\begin{document}

\maketitle

\vspace{-20pt}
\begin{center}
\noindent\textsuperscript{1} Department of Mathematical Sciences, The University of Texas at El Paso. \\
\textsuperscript{2} Departments of Neurosurgery and Biomedical Informatics Research, Stanford University. \\~\\
\end{center}

\begin{abstract}
{\normalsize Competing risk analysis accounts for multiple mutually exclusive events, improving risk estimation over traditional survival analysis. Despite methodological advancements, a comprehensive comparison of competing risk methods, especially in high-dimensional settings, remains limited. This study evaluates penalized regression (LASSO, SCAD, MCP), boosting (CoxBoost, CB), random forest (RF), and deep learning (DeepHit, DH) methods for competing risk analysis through extensive simulations, assessing variable selection, estimation accuracy, discrimination, and calibration under diverse data conditions. Our results show that, under the considered settings, CB provides strong control of false discoveries, stable estimation, and competitive discriminative ability, particularly in high-dimensional settings, while MCP and SCAD provide improved calibration in $n>p$ scenarios. RF and DH are effective at capturing nonlinear effects, but in the present implementation, they tend to exhibit weaker performance, with RF identifying broader variable sets and DH showing limited calibration accuracy. We further illustrate the application of these methods through an analysis of a melanoma gene expression dataset with survival outcomes. This study provides comparative evidence and preliminary guidelines for selecting competing risk models in high-dimensional settings and outlines important directions for future research.

}
\end{abstract}
\vspace{-2em}  % Adjust vertical spacing to bring the Introduction closer
\noindent

\twocolumn

%☀️☀️☀️
%☀️☀️☀️
\section{Introduction} \label{sec:introduction}
Competing risk analysis is an extension of survival analysis for settings where multiple mutually exclusive events can occur, such that the occurrence of one event precludes the others. Traditional survival methods assume a single event type and typically treat competing events as independent censoring, implying that if one event occurs, the other could still happen given more time. While this assumption is valid for censoring due to study termination or withdrawal, it fails in the presence of competing risks, where one outcome inherently precludes another \citep{beyersmann_et_al_2012, pintilie_2006}.

For example, in transplantation studies, competing risks include death from the underlying disease or from transplant-related complications: once one occurs, the other cannot. In such cases, standard survival methods like the Kaplan–Meier estimator can overestimate survival probabilities, since they ignore the impact of competing events \citep{kalbfleisch_prentice_2002}. Likewise, the Cox proportional hazards model assumes proportional hazards across event types, which may not hold in competing risk settings \citep{fine_gray_1999}.

Competing risk analysis explicitly incorporates alternative outcomes, yielding more accurate and interpretable inferences. This is particularly critical in healthcare, where patients may die from different causes (e.g., cancer vs. cardiovascular disease) \citep{andersen_et_al_2012}, and in engineering, where systems may fail due to distinct mechanisms \citep{crowder_2001}.

The most widely used statistical approaches are proportional cause-specific hazards (PCSH) regression \citep{prentice_et_al_1978}, which models cause-specific hazard functions by treating competing events as censored, and proportional subdistribution hazards (PSDH) regression \citep{fine_gray_1999}, also known as the Fine–Gray model, which models the cumulative incidence function. A third approach models the joint distribution of event time and type \citep{larson_dinse_1985,ng_mclachlan_2003,nicolaie_et_al_2010}. Several comparative reviews of these methods are available \citep{haller_et_al_2013, bakoyannis_touloumi_2012, zhang_et_al_2008}.

However, these traditional methods are not directly applicable to high-dimensional data, where the number of covariates may exceed the sample size. For variable selection, estimation, and prediction in such settings, a number of machine-learning approaches have been proposed. Recently, \cite{monterrubio_et_al_2024} provided a broad methodological review of various statistical and machine learning approaches for competing risk analysis. However, they did not conduct simulation analyses to evaluate practical performance and applied these methods only to low-dimensional real-world data. Similarly, \cite{kantidakis_et_al_2023} used a low-dimensional extremity soft-tissue sarcoma dataset to compare the performance of PCSH and PSDH models alongside three machine learning approaches---a partial logistic artificial neural network for competing risks (PLANNCR) \citep{biganzoli_et_al_2006}, an extended PLANNCR with novel architectural specifications \citep{lisboa_et_al_2009}, and a random survival forest for competing risks \citep{ishwaran_et_al_2014}.

Comparative studies that incorporate high-dimensional competing risk data are more limited. \cite{tapak_et_al_2015} compared three variable selection techniques---LASSO, elastic net, and likelihood-based boosting---using a simulation study with 5,000 covariates and 400 observations and a bladder cancer dataset containing 1,386 clinical and microarray features for 301 patients. \cite{hou_et_al_2018} compared LASSO (under PCSH and PSDH), adaptive LASSO (under PCSH and PSDH), and likelihood-based boosting (under PSDH) using simulated datasets with sample size 500 and number of covariates 20, 500, and 1000, independent, exchangeable, and AR(1) correlation structures for continuous covariates, and balanced and sparse binary covariates and a prostate cancer data from the SEER-Medicare linked dataset containing 57,011 patients and 8,984 clinical, demographical, and insurance claim covariates.

In this study, we conduct an extensive simulation-based comparison of competing risk methods in a broader set of high-dimensional scenarios than considered in previous works, using data generated from the PSDH model. We comparatively evaluate penalized regression (PR) approaches using LASSO, SCAD, and MCP penalties \citep{fu_et_al_2017}, likelihood-based boosting (CoxBoost, CB) \citep{binder_et_al_2009}, random forest (RF) \citep{ishwaran_et_al_2014}, and deep learning (DeepHit, DH) \cite{lee_et_al_2018}. While \cite{ishwaran_et_al_2014} compared RF and CB, their study was conducted in a less diverse setting than ours. Additionally, previous works have not compared PR methods of \citep{fu_et_al_2017} with CB and RF in a simulation framework. Likewise, the DH method has not been compared with PR, RF, or CB models in competing risk analysis.

To address this gap in comparative studies, we systematically evaluate and compare PR, CB, RF, and DH methods using simulated datasets across a broad spectrum of data-generating conditions, including sample sizes, number of covariates, correlation structures among continuous covariates, sparsity levels in discrete covariates, and different covariate effect models. Since the simulations are based on the Fine--Gray (PSDH) model, methods aligned with this framework (such as PR and CB) may be relatively advantaged; accordingly, the study should be interpreted as a structured comparison within a widely used modeling paradigm rather than a fully model-agnostic benchmark. All methods are implemented using standard procedures and tuning strategies as provided in their respective packages and original references. Based on our results, we provide practical guidelines for selecting competing risk models, while noting in the Discussion directions for extending the comparison to alternative generative settings to assess robustness. Additionally, we illustrate the comparative utility and flexibility of these methods through the analysis of a high-dimensional melanoma gene expression dataset, which includes survival outcomes for 214 patients and over 47,000 gene expressions.

The remainder of this paper is organized as follows: Section~\ref{sec:background} provides a formal background on competing risk analysis, Section~\ref{sec:methods} describes the competing risk methods considered in this study, Section~\ref{sec:simulation} presents the simulation study, detailing data generation, evaluation metrics, and results, and Section~\ref{sec:realdata} reports the melanoma gene expression data analysis. Finally, Section~\ref{sec:conclusion} provides an overall discussion and concluding remarks.

%☀️☀️
\subsection{Competing Risk Definition}
There are two distinct approaches to defining competing risk data: one based on a bivariate random variable, extending the definition of ordinary survival data, and the other using latent failure times \citep{beyersmann_et_al_2012, pintilie_2006}.

\begin{itemize}
\item \emph{The Bivariate Random Variable Approach} is the more commonly used approach, which models the competing risk data for an individual, who can experience $K \geq 2$ competing events or be censored, using a pair of random variables $(T,E)$, where 
\thickmuskip=0.5\thickmuskip
$$
E = 
\begin{cases}
0, & \text{individual is censored} \\
k, & \text{individual experiences event } k, \; k = 1, \ldots, K,
\end{cases}
$$
is the event type indicator and
$$
T = 
\begin{cases}
\text{time to censoring}, & \text{if } E = 0 \\
\text{time to event } k, & \text{if } E = k,
\end{cases}
$$
is time-to-event, where both $T$ and $E$ are stricltly positive random variables.

\item \emph{The Latent failure Time Approach} defines the competing risk data using a set of independent latent times \((T_1, \ldots, T_K)\) for \(K \geq 2\) competing events, where each $T_k$ represents the unobserved time of event $k$. In this framework, for uncensored individuals, only the minimum time $T = \min\{ T_1, \ldots, T_K \}$ is observed and for censored individuals, $T$ is the time to censoring. 
\end{itemize}

Despite the utility of the latent failure time approach, \cite{prentice_et_al_1978} raised concerns that its correlation structure cannot be estimated from observed data, making inference problematic. Consequenty, they discouraged its application in competing risk analysis. \citep{pintilie_2006, beyersmann_et_al_2012} demonstrated that the bivariate random variable approach more effectively captures the relationship between event type and time to occurrence compared to the latent failure time framework. In the remainder of this work, we adopt the bivariate random variable approach, ensuring that all definitions and analyses align with this framework.

%☀️☀️
\subsection{Important Measures for Competing Risk Data}
In the context of competing risks, there are measures that extend ordinary survival analysis measures to multiple events for accurately assessing the probability of each competing event. These measures allow to analyze the risk dynamics taking into account the particularities of each competing event to avoid potential biases in the estimation of survival or time to a specific event. We describe these measures below.

%☀️
\subsubsection*{Cumulative Incidence (CI):}
In competing risk analysis, the cumulative incidence function, $F_k(t)$, represents the probability of experiencing event $k$ by time $t$ and is defined as:
$$
F_k(t) = P(T \leq t, E = k),
$$
The total cumulative incidence function, $F(t)$, is the sum of the CIs for all competing events, $F(t) = \sum_{k=1}^K F_k(t)$.

The overall survival function, $S(t)$, represents the probability that no event has occurred by time $t$ and is given by: 
$$
S(t) = P(T > t) = 1 - \sum_{k=1}^K F_k(t).
$$
This formulation ensures that survival probability properly accounts for competing risks, avoiding overestimation common in traditional survival analysis.
%☀️
\subsubsection*{Cause-specific Event Density:} The cause-specific event density, $f_k(t)$, represents the instantaneous rate at which events of type $k$ occur at time $t$. It is defined as:
$$
f_k(t) = F'_k(t) = \lim_{\Delta t \to 0} \frac{P(t \leq T \leq t + \Delta t, E = k)}{\Delta t},
$$
where $\Delta t$ denotes an infinitesimal time and $f_k(t)$ exists only if $F_k(t)$ is differentiable.

%☀️
\subsubsection*{Cause-specific Hazard (CSH):}
The cause-specific hazard function, $h_k^{CSH}(t)$, represents the instantaneous rate at which events of type $k$ occur at time $t$, given that no event has yet occurred. It is defined as
$$
 h_k^{CSH}(t) = \lim_{\Delta t \to 0} \frac{P(t \leq T \leq t + \Delta t, E = k \mid T \geq t)}{\Delta t} = \frac{f_k(t)}{S(t-)},
$$
where $S(t-)$ is the left-continuous survival function. In other words, CSH answers the question: `If a person has survived event-free up to time $t$, how likely are they to have event type $k$ right now?' This makes it especially useful when the scientific focus is on the direct biological or clinical process leading to an event.

The overall hazard function, $h(t)$, is obtained by summing up the cause-specific hazards of all competing events:
$$
h(t) = \sum_{k=1}^{K} h_k^{CSH}(t).
$$

The cumulative cause-specific hazard function, $H_k^{CSH}(t)$, represents the total hazard exposure for event $k$ up to time $t$ and is given by:
$$
H_k^{CSH}(t) = \int_{0}^{t} h_k^{CSH}(s) \, ds,
$$
which reflects the total exposure to the hazard of event $k$ up to time $t$.

Similarly, the overall cumulative hazard function, $H(t)$, is:
$$
H(t) = \sum_{k=1}^{K} H_k^{CSH}(t),
$$ 
which is related to the survival function, $S(t)$, through
$$
S(t) = e^{-H(t)}.
$$

It is important to note that the relationship between the CSH $h_k(t)$ and CI $F_k^{CSH}(t)$ is not straightforward. Specifically, $F_k^{CSH}(t)$ is not simply given by $1 - e^{-H_k^{CSH}(t)}$, as it depends on the CSHs of all competing events through the relationship
$$
F_k^{CSH}(t) = \int_{0}^{t} h_k^{CSH}(s) e^{-\sum_{k=1}^{K} H_k^{CSH}(s)} \, ds.
%\int_{0}^{t} f_k(s) \, ds = \int_{0}^{t} h_k^{CSH}(s) S(s-) \, ds \\
$$

%☀️
\subsubsection*{Subdistribution Hazard (SDH):}
The subdistribution hazard function, $h_k^{SDH}(t)$, for event $k$ at time $t$ represents the instantaneous rate of occurrence of event $k$ at time $t$, given that event $k$ has not occurred by time $t$ (but event $k'\ne k$ may have occurred by time $t$). It is given as:
{\small
$$
\lim_{\Delta t \to 0} \frac{P(t \leq T \leq t + \Delta t, E = k \mid T \geq t \cup \{T < t, E \neq k\})}{\Delta t}.
$$
}
SDH answers the question: `In the overall study population, including those who may have already had other types of events, how likely are we to observe event type $k$ at time $t$?' Thus, while CSH for event $k$ treats competing events as censored, SDH for event $k$ explicitly accounts for the presence of competing risks. This makes SDH especially valuable for predicting cumulative incidence in practice.

Unlike CSH, the sum of SDHs of all competing events does not equal the overall hazard:
$$
h(t) \neq \sum_{k=1}^{K} h_k^{SDH}(t).
$$
The cumulative subdistribution hazard, $H_k^{SDH}(t)$, for event $k$ at time $t$ is given by:
$$
H_k^{SDH}(t) = \int_0^t h_k^{SDH}(s) ds.
$$
There is a direct relationship between the SDH, $h_k^{SDH}(t)$, and CI,  $F_k(t)$, for event $k$:
$$
F_k^{SDH}(t) = 1 - e^{-H_k^{SDH}(t)}.
$$

%☀️
\subsubsection*{Relationship Between CSH and SDH:}
The relationship between the CSH and SDH can be derived analytically through their connection the CI. Specifically, for two competing events \citep{beyersmann_et_al_2012}:
$$
h_1^{CSH}(t) = h_1^{SDH}(t) \left( 1 + \frac{F_2^{CSH}(t)}{S(t)} \right),
$$
where $h_1^{CSH}(t)$ and $h_1^{SDH}(t)$ are respectively the CSH and the SDH for event 1, $F_2^{CSH}(t)$ the CI for event 2, and $S(t)$ the corresponding overall function at time $t$.

%☀️☀️
\subsection{Most Used Competing Risk Models}

%☀️
\subsubsection*{Proportional CSH (PCSH) Model:} The PCSH model is a semi-parametric regression model that extends the Cox proportional hazards model to competing risk data. It estimates the effect of covariates on CSH under the assumption that CSHs of different individuals are proportional. The PCSH model is given by \citep{prentice_et_al_1978}
$$
h_k^{CSH}(t|\bfx) = h_{k,0}(t) \exp({\bfbeta_k^{CSH}}^T\bfx),
$$
where $h_{k,0}(t)$ is a nonparametric baseline CSH for event $k$, $\bfx$ is the vector of covariates, and $\bfbeta_k^{CSH}$ is the vector of covariate effects for event $k$.

%☀️
\subsubsection*{Proportional SDH (PSDH) model:} 
The PSDH model, or the Fine–Gray model, is a semi-parametric regression model that estimates the effect of covariates on SDH under the assumption that SDHs of different individuals are proportional to each other. The PSDH model is given by \citep{fine_gray_1999}
$$
h_1^{SDH}(t|\bfx) = h_{1,0}(t) \exp({\bfbeta_1^{SDH}}^T\bfx),
$$
where $h_1^{SDH}(t|\bfx)$ is the SDH for the event of interest (WLG indexed as event $k=1$), $h_{1,0}(t)$ is a nonparametric baseline SDH for the event of interest, $\bfx$ is the vector of covariates, and $\bfbeta_1^{SDH}$ is the vector of covariate effects for the event of interest. Two important points to note are that:
\begin{itemize}
\item Since $h_1^{SDH}(t)$ and $F_1^{SDH}(t)$ are directly related, the Fine–Gray model directly models the effect of covariates on the CI, $F_1^{SDH}(t)$.
\item The Fine–Gray model is expressed in terms of a single event of interest rather than all competing events. This is because if separate Fine–Gray models are fit for each event, the estimated covariate effects for one event inherently depend on the competing events. This happens because the SDH accounts for individuals who have either not yet experienced any event or have already experienced a competing event. Hence, the proportionality assumption may not hold consistently across all competing events, as the risk of one event dynamically depends on the occurrence of others.
\end{itemize}

%☀️☀️☀️
\section{Modern Competing Risk Analysis Methods} \label{sec:methods}
In this section, we provide an introduction to the modern competing risk analysis methods that are compared in this study (see Section~\ref{sec:introduction})---namely, PR, CB, RF, and DH---through extensive simulation analysis. Of these, CB, RF, and DH can tackle high-dimensional data, whereas PR is not designed for high-dimensional data. All the methods consider right-censored survival data.

Henceforth, we are going to adopt the following notation consistently across all methods. Suppose, we have right-censored survival data for $n$ subjects. For each subject $i$, $i \in \{1,2,\ldots, n\}$, the data consists of $(T_i,\delta_i,\epsilon_i,\bfx_i)$, where $T_i = T^*_i \wedge C_i$ is the observed time, where $T^*_i$ is true event time (time to first event occurrence) and $C_i$ is the censoring time, independent of $T^*_i$, $\delta_i = I(T^*_i \le C_i)$ is the censoring indicator ($\delta_i=0$ if subject $i$ is censored, or 1 otherwise), $\epsilon_i \in \{1,2,\ldots,K\}$ is the event type indicator (not specified when $\delta_i=0$), and $\bfx_i^{p\times 1}$ is the set of covariate values, $p$ being the number of covariates. Further, suppose $\delta_i^* = \delta_i\epsilon_i$ so that $\delta_i\epsilon_i \in \{0,1,2,\ldots,K\}$.

%☀️☀️
\subsection{Penalized Regression} \label{sec:PR}
Penalized regression is a powerful approach used to improve model accuracy and perform variable selection by introducing a penalty term to the model. This penalty is either added to the loss function to be minimized (such as the residual sum of squares in linear models) or subtracted from the likelihood function in more complex models like logistic or Cox regression. Penalized regression improves the accuracy of estimates by shrinking the coefficients of non-influential covariates to zero, thus selecting only the relevant covariates. This helps avoid overfitting when dealing with large datasets with many covariates. As a result, the model becomes less specific to the original dataset and performs better on new data \citep{hastie_et_al_2009}. 

In the context of competing risk data, we consider the penalized regression method proposed by \cite{fu_et_al_2017} for the PSDH model \citep{fine_gray_1999}. This method enables simultaneous variable selection and parameter estimation by maximizing a penalized log-partial likelihood function, given, in accordance with our notation, by:   
%The PSH model is used to assess the cumulative incidence function (CIF), which quantifies the probability of failure from a particular cause in the presence of competing risks. In this framework, the goal is to identify which covariates significantly impact the risk of an event occurring while taking into account other potential outcomes that prevent the event of interest occurring \citep{}.
$$
Q\left(\bfbeta_1^{SDH}\right) = l\left(\bfbeta_1^{SDH}\right) - n \sum_{j=1}^{p} p_{\lambda}\left(\left|\beta_{1,j}^{SDH}\right|\right),
$$ 
where
\thickmuskip=0.1\thickmuskip
$$
\begin{aligned}
& l\left(\bfbeta_1^{SDH}\right) = \sum_{i=1}^{n} \int_0^{\infty} \left[ {\bfbeta_1^{SDH}}^T \bfx_i \right. \\
& \left. - \log \left( \sum_{j} w_j(u) Y_j(u) \exp\left({\bfbeta_1^{SDH}}^T \bfx_j\right) \right) \right] w_i(u) \, dN_i(u)
\end{aligned}
$$
\thickmuskip=10\thickmuskip
is the log-partial likelihood function of the PSDH model \citep{fu_et_al_2017}, $w_i(t)=I(C_i\ge T_i \wedge t)\hat{G}(t) / \hat{G}(T_i\wedge t)$ represents time-dependent inverse probability of censoring weights, where $G(t)=Pr(C>t)$ is the survival function of the censoring variable and $\hat{G}(t)$ is its Kaplan-Meier estimator, $Y_i(t)$ is the at-risk process, which is 1 if individual $i$ is at risk at time $t$, and 0 otherwise, $N_i(t)$ is the counting process that tracks the occurrence of the event of interest for individual $i$ by time $t$, $p_{\lambda}(.)$ is the penalty function, where $\lambda$ is a tuning parameter controling model complexity, and $\beta_{1,j}^{SDH}$ is the $j$th component of $\bfbeta_1^{SDH}$.

Within this framework, we explore three distinct penalty functions: LASSO, SCAD, and MCP. Whereas LASSO is a convex penalty, SCAD and MCP are non-convex penalties. Each offers a unique approach to balancing model complexity and accuracy. The differences in how each penalty function behaves are highlighted below.
%particularly in how SCAD and MCP adjust with varying values of the tuning parameter \(a\).

\begin{itemize}
\item LASSO: The LASSO penalty is defined as
$$ 
p_{\lambda}^{LASSO}\left(\left|\beta_{1,j}^{SDH}\right|\right) = \lambda \left|\beta_{1,j}^{SDH}\right|,
$$
which penalizes coefficients proportionally to their absolute values \citep{tibshirani1997lasso}, leading to shrinkage and sparsity. Since LASSO applies equal shrinkage to all coefficients, it may over-penalize larger ones, biasing important variables toward zero. In contrast, SCAD and MCP relax the penalty on large coefficients, reducing excessive shrinkage \citep{fu_et_al_2017}.

\item SCAD: The SCAD penalty is expressed in terms of its first derivative \citep{fu_et_al_2017, fan_li_2001}:
\begin{align*} 
&{p_{\lambda}^{SCAD}}'\left(\left|\beta_{1,j}^{SDH}\right|\right) = \lambda I\left(\beta_{1,j}^{SDH} \leq \lambda\right) \; +  \\
&\hspace{2.5cm}\frac{\left(a\lambda - \left|\beta_{1,j}^{SDH}\right|\right)_+}{(a - 1)\lambda} I\left(\beta_{1,j}^{SDH} > \lambda\right),
\end{align*}
where $a > 2$ is a tuning parameter that determines when large coefficients stop being penalized. For small coefficients ($|\beta_j| \leq \lambda$), SCAD behaves like LASSO; for intermediate ones ($\lambda < |\beta_j| \leq a\lambda$), the penalty decreases smoothly; and for large ones ($|\beta_j| > a\lambda$), no penalty is applied. This adaptive behavior prevents over-penalization and preserves moderate-to-large predictors. 

\item MCP: The MCP penalty is also expressed using its first derivative \citep{fu_et_al_2017, zhang_2010}:
$$
\begin{aligned}    
&{p_{\lambda}^{MCP}}'\left(\left|\beta_{1,j}^{SDH}\right|\right) = \lambda I\left(\left|\beta_{1,j}^{SDH}\right| \leq a\lambda\right) \, \text{sign}(\beta_{1,j}^{SDH}) \\
& \hspace{4cm} \left(\lambda - \frac{\left|\beta_{1,j}^{SDH}\right|}{a}\right),
\end{aligned}
$$
where $a > 1$ controls the decay rate of the penalty for large coefficients. Unlike SCAD, MCP has no constant penalization region and immediately reduces the penalty as $|\beta_{1,j}^{SDH}|$ increases. This design lessens shrinkage for large coefficients, improving variable selection while retaining sparsity.
\end{itemize}

%☀️☀️
\subsection{Boosting}
Boosting is an iterative machine learning ensemble technique that combines multiple weak learners, with each iteration correcting the errors of its predecessor to build a stronger, more accurate model. It dynamically adjusts data weights, emphasizing misclassified observations, thereby improving prediction accuracy while reducing errors and overfitting \citep{hastie_et_al_2009}. Boosting is particularly effective in sparse high-dimensional scenarios, where only a small subset of covariates significantly influences the response. This makes it especially useful in genomic studies, such as gene expression analysis or SNP selection in GWAS.
%Boosting is a method in machine learning that combines several less accurate models or weak learning algorithms to create a more strong and accurate overall model. It works by training a series of simple models, called weak learners, where each model focuses on correcting the mistakes of the previous one. In this process, the algorithm updates the weights of the training data after each iteration. Data points that were misclassified by the previous model are given more weight, so that in the next model emphasizes on correcting those mistakes.This iterative process refines the model by focusing on correcting the errors made by previous models, which improves the overall accuracy. By doing this, the combined model gets better at making accurate predictions. Boosting helps reduce prediction errors and prevents the model from becoming too specific to the training data in many cases.

\cite{binder_et_al_2009} proposed a componentwise, likelihood-based boosting method for fitting the PSDH model in high-dimensional data. This method differentiates between mandatory covariates, which must be included in the model, and optional covariates, which may or may not be included. The method iteratively minimizes a loss function, updating covariate effects step by step. Before each boosting step, mandatory covariates are updated simultaneously. Then, in each step, one optional covariate is selected and updated based on minimizing a penalized partial likelihood function, where the previous boosting steps are incorporated through an offset term. 

Let $\mathcal{J}_{mand}$ and $\mathcal{J}_{opt}$ respectively denote the indices for mandatory and optional covariates, such that $\mathcal{J}_{mand} \cup \mathcal{J}_{opt} = \{1,\ldots,p\}$. The boosting algorithm follows these steps:

\begin{outline}
\1 \tb{Initialization}: Set the initial offset $\eta_{(0)i} = 0$ for all $i=1,\ldots,n$ and initialize the parameter vector $\bfbeta_{1(0)}^{SDH} = (0, \dots, 0)'$.

\1 \textbf{Boosting Steps}: For each boosting step $m=1,\dots,M$, 

\2 Update parameters for mandatory covariates, $\beta_{1(m-1),j}^{SDH}, j \in \mathcal{J}_{mand}$, by one maximum partial likelihood Newton--Raphson step and then update the offset as $\widehat{\eta}_{(m-1)i} = \bfx_i^T \widehat{\bfbeta}_{1(m-1)}^{SDH}$.

\2 For each $j \in \mathcal{J}_{opt}$, estimate the parameters $\gamma_{(m)j}$ in candidate models
$$
\begin{aligned}
&h_1^{SDH}(t|x_i) = h_{1,0}(t) \exp(\widehat{\eta}_{(m-1)i} + \gamma_{(m)j} x_{ij}),\\
& \hspace{5.5cm} i=1,\ldots,n,
\end{aligned}
$$
as 
$$
\widehat{\gamma}_{(m)j} = I_{pen}^{-1}(0) U_{pen}(0),
$$ 
where $U_{pen}(\gamma)=\partial l_{pen}(\gamma)/\partial\gamma$ is the score function, $I_{pen}(\gamma)=\partial^2 l_{pen}(\gamma)/\partial^2\gamma$ is the information matrix, and $l_{pen}(\gamma)$ is the penalized log-likelihood given by
$$
\begin{aligned}
&l_{\text{pen}}(\gamma_{(m)j}) = \sum_{i=1}^{n} I(\delta_i^* = 1) \left( \hat{\eta}_{(m-1)i} + \gamma_{(m)j} x_{ij} \right) \\
&- \log\left( \sum_{l \in R_i} w_l(t_i) \exp\left(\hat{\eta}_{(m-1)i} + \gamma_{(m)j} x_{lj} \right) \right) \\
&+ \frac{\lambda}{2} \gamma_{(m)j}^2,
\end{aligned}
$$
where $w_i(t)$ is the time-dependent weights based on inverse probability of censoring defined earlier, and $R_i$ is the risk set at time $T_i$ for subject $i$, and $\lambda$ is a penalty parameter, determining the size of the boosting steps.
	 
\2 Select the best candidate $j^*$ as the one that maximizes the score statistic $U'_{pen}(0) I_{pen}^{-1}(0) U'_{pen}(0)$. Then, update the parameter vector $\bfbeta_{1(m-1)}^{SDH}$ as follows:
$$
\widehat{\beta}_{1(m),j}^{SDH} =
\begin{cases} 
\widehat{\beta}_{1(m-1),j}^{SDH} + \widehat{\gamma}_{(m)j^*}, & \text{if } j = j^* \\
\widehat{\beta}_{1(m-1),j}^{SDH}, & \text{otherwise}.
\end{cases}
$$

where $\widehat{\beta}_{1(m),j}^{SDH}$ is the parameter estimate for covariate $j$ at step $m$ and $\widehat{\beta}_{1(m-1),j}^{SDH}$ is the parameter estimate from the previous boosting step, serving as the baseline for updates.
\end{outline}

After obtaining the boosting fit, all covariates with non-zero estimates are considered selected.

%The objective is to minimize a penalized likelihood function at each boosting step for the optional covariates, which helps balance how well the model fits the data with its complexity.

%The algorithm operates by iteratively refining the subdistribution hazard function, systematically selecting and updating the most influential covariates at each step. By doing so, it constructs a sparse model, where only the most relevant covariates contribute to the final predictions, ensuring a more interpretable and efficient outcome. This approach is highly effective in handling competing risk scenarios, where traditional models may struggle with large datasets and multiple covariates \citep{binder_et_al_2009}.

%☀️☀️
\subsection{Random Forest}
Random Forest is an ensemble learning method that constructs multiple decision trees using bootstrapped samples of the training data and randomly selected feature subsets at each split. This randomness reduces  overfitting while improving model generalization. Additionally, the method utilizes out-of-bag (OOB) data--samples not used in tree construction--to estimate model accuracy and assess variable importance. For classification tasks, predictions are made based on majority voting among the trees, while in regression, the final output is the average prediction across trees. Due to its ability to handle high-dimensional data, manage missing values, and provide feature importance insights, RF is widely applied across various domains \citep{breiman_2001}.

\citep{ishwaran_et_al_2014} introduced Random Survival Forests for competing risk data, designed to model non-linear effects and interactions, making it particularly effective for high-dimensional settings. The competing risk forest follows the same fundamental structure as standard RF but differs in the splitting rule and prediction measures computed in the leaves. The algorithm is described below:
\begin{outline}
\1 Draw $B$ bootstrap samples from the learning data.

\1 For each bootstrap sample, grow a competing risk tree as follows. At each node, randomly select $M(\le p)$ candidate variables and split the node using the variable that maximizes a competing risk splitting rule. Two splitting rules are considered, which are given below. \\

Let $t_1<t_2<\ldots<t_m$ denote the distinct observed event times,\\
$N_k(t)$ = number of type $k$ events in $[0,t]$\\
$N(t)$ = the total number of events in $[0,t]$, \\
$Y(t)$ = number of individuals at risk (event-free or uncensored) just before $t$, \\
$d_k(t_l)$ = number of type $k$ events at $t_l$ = $\sum_{i=1}^n I(T_i=t_l, \epsilon_i=k)$, and\\
$d(t_l)$ = total number of events at $t_l$ = $\sum_k d_k(t_l)$. \\

The Kaplan-Meier estimator of event-free survival function $S(t)$ is: 
$$
\widehat{S}(t) = \sum_{l=1}^{m(t)} \left( 1 - \frac{d(t_l)}{Y(t_l)} \right)
$$
where $m(t) = \max\{l: t_l\le t \}$.
The Aalen-Johansen estimator of CI is:
$$
\widehat{F}_k(t) = \sum_{l=1}^{m(t)} \widehat{S}(t_{l-1}) \frac{d_k(t_l)}{Y(t_l)} 
$$

Now, assume a node is split based on a continuous covariate $x$ into a left (le, where $x\le c$) and right (ri, where $x>c$) daughter node for a scalar $c$. Let $t_{m_{le}}$ and $t_{m_{ri}}$ denote the largest observed event time in the left and right daughter nodes, respectively. Equivalently to the notations used for the unpartitioned data above, for the left and right partitions, consider the notations\\
$Y_{le}(t) = \sum_{i=1}^n I(T_i\ge t, x_i\le c)$, \\
$Y_{ri}(t) = \sum_{i=1}^n I(T_i\ge t, x_i> c)$, \\
$Y(t) = Y_{le}(t) + Y_{ri}(t)$,\\
$d_{k,le}(t) = \sum_{i=1}^n I(T_i=t, \epsilon_i=k, x_i\le c)$, \\
$d_{k,ri}(t) = \sum_{i=1}^n I(T_i=t, \epsilon_i=k, x_i> c)$, \\
$d_k(t) = d_{k,le}(t) + d_{k,ri}(t)$. \\
Further, let $\widehat{F}_{k,le}(t)$ and $\widehat{F}_{k,ri}(t)$ denote the Aalen-Johansen estimator of CI for the left and right partitions, respectively.\\

Using these notations, the splitting rules are based on:

\2 \tb{Log-rank test:} This test compares the cause-specific hazard between the two daughter nodes, $H_0:h_{k,le}^{CSH}(t)=h_{k,ri}^{CSH}(t)$ for all $t\le \tau,$ the largest observed time. For competing event $k$, the test statistic is
$$
\begin{aligned}
L_k^{LR}(x,c) =& \frac{1}{\hat{\sigma}_k^{LR}(x,c)} \sum_{l=1}^m W_k(t_l) \\
&\left( d_{k,le}(t_l) - \frac{d_k(t_l)Y_{le}(t_l)}{Y(t_l)} \right),
\end{aligned}
$$
where
$$
\begin{aligned}
(\hat{\sigma}_k^{LR}(x,c))^2 &= \sum_{l=1}^m W_k(t_l)^2 d_k(t_l) \frac{Y_{le}(t_l)}{Y(t_l)} \\
&\left(1-\frac{Y_{le}(t_l)}{Y(t_l)}\right) \left( \frac{Y(t_l)-d_k(t_l)}{Y(t_l)-1} \right),
\end{aligned}
$$
and $W_k(t)>0$ are time-dependent weights. $W_k=1$ yields the standard log-rank test, optimal for detecting proportional CSHs. The optimal split maximizes $\left|L_k^{LR}(x,c)\right|$.

\2 \tb{Gray’s test:} This test compares the cause-specific cumulative incidence function between the two daughter nodes, $H_0:F_{k,le}(t)=F_{k,ri}(t)$ for all $t\le \tau$. Assuming $k=1$ and $K=2$ (pooling all events except event 1), the test statistic is 
$$
\int_0^{t_l} W_k(s)V_{le}(s) \left\{ \frac{d\widehat{F}_{k,le}(s)}{1-\widehat{F}_{k,le}(s)} - \frac{\widehat{F}_k(ds)}{1-\widehat{F}_k(s)} \right\},
$$
where
$$
\begin{aligned}
V_{le}(t) =& I\{t_{m_{le}} \ge t\} Y_{le}(t) [1 - \widehat{F}_{k,le}(t-)] \\
&(\widehat{S}_{le}(t-))^{-1}.
\end{aligned}
$$
For the special case where the censoring is due purely to administrative loss to follow-up time, meaning no subjects are lost to follow-up unpredictably (e.g., dropping out of the study), the potential censoring time is known (to be the study's administrative cutoff) for those subjects who experiences an event before the end of follow-up. For such case, the score statistic, $L_k^G(x,c)$, of Gray's test is obtained by substituting the modified risk set $Y_k^*(t) = \sum_{i=1}^n I(T_i\ge t \cup (T_i<t \cap \delta_i\epsilon_i\ne j \cap C_i > t))$ for $Y(t)$ and similarly defined $Y_{k,le}^*(t)$ for $Y_{k,le}(t)$ in the expression for $L_k^{LR}(x,c)$. Then, $|L_k^G(x,c)|$ is maximixed over $(x,c)$ pair to obtain the optimal split. Even when the potential censoring time, indicated above, is not known, using the largest observed time provides a good approximation of $L_k^G(x,c)$ as per the authors.

\1 Grow each tree so long as the number of unique cases in a terminal node does not go below $n_0(>0)$.

\1 For each tree, $b$, $b=1,2,\ldots,B$, compute the estimates $\left(\widehat{F}_{k,b}, \widehat{H}_{k,b}^{CSH}, \widehat{M}_{k,b}(\tau)\right)_{k=1,2,\ldots,K}$, $\widehat{S}_b$, and $\widehat{H}_b$.

\1 Take the average of each estimate over the $B$ trees to obtain its ensemble estimate.
\end{outline}

The above algorithm can also be used to compute variable importance and minimal depth for each covariate, which aids in variable selection \citep{ishwaran_et_al_2014}.

%☀️☀️
\subsection{Deep Learning}

%Deep learning is rooted in networks, a class of computer models designed to imitate the structure and workings of the brain. A neural network is made up of interconnected units called referred to as neurons that are organized into layers. Each layer processes input data using a sequence of transformation operations to enable the model to adjust its weights and bias parameters, and to learn and capture complex features.During the training process, the adaptive weighting and bias parameters of networks are adjusted to reduce errors and enhance their ability to understand models better. The hierarchical organization of neural networks enables them to capture relationships, in data effectively. This makes neural networks a valuable tool, for recognizing patterns and engaging in analysis \citep{bishop2006pattern}. Deep learning uses layered structures to grasp complex ideas by blending simpler ones in a hierarchical manner. Each layer within the model captures different levels of complexity and enables the network to automatically learn and represent complicated patterns from data . This layered approach reduces the need for manual feature engineering and allows models to adapt across applications like categorizing images or recognizing speech and language \citep{goodfellow2016deep}. 

Deep learning is based on neural networks, computational models inspired by the brain’s structure. These networks consist of interconnected units, called neurons, arranged in layers, each processing input data through a sequence of transformations enabling the model to adjust its weights and biases for learning complex features. During training, these parameters adapt to minimize errors and enhance pattern recognition. The hierarchical structure enables automatic feature extraction, reducing the need for manual engineering and making deep learning effective for tasks like image classification and speech recognition \citep{bishop2006pattern, goodfellow2016deep}.

%The proposed method addresses the limitations of traditional survival analysis models in the context of competing risks. Traditional survival models, like the Cox proportional risk model tend to rely on parametric assumptions that may not always hold true in practical real world datasets. Traditional models also have challenges to manage the complexity of competing risks where multiple types of events could potentially take place. DeepHit stands out from traditional approaches by capturing the combined distribution of survival time and event types through a deep neural network model.   

The DH model, proposed by \citep{lee_et_al_2018}, is a deep learning framework designed to handle survival analysis with competing risks. Unlike traditional survival models, this method does not assume a specific form for the time-to-event distribution. Rather it constructs a deep neural network to learn both the form and the parameters of the distribution directly from the data, and can learn potentially non-linear and/or non-proportional relationships between covariates and risks. Using our notation, the output of the model is the estimated joint probability of survival time and competing event type, $y_{\delta^*,t}=\widehat{Pr}(t, \delta^* | \bfx)$, which denotes the probability that a (new) individual with covariates $\bfx$ will experience an event of type $\delta^*$ (including censoring) at a specific time $t$. 

DH's architecture includes a multitask network consisting of a shared sub-network and $K$ cause-specific sub-networks to learn shared and cause-specific representations, respectively. A single softmax layer is used as the output layer that learns the aforesaid joint distribution. The shared subnetwork consists of $L_T$ fully connected layers, accepts as input the covariates $\bfx$, and outputs a vector $f_t(x)$ that captures the common (latent) representation of the competing events. The $k$-th cause-specific subnetwork consists of $L_{C,k}$ fully connected layers, accepts as input the pair $\bfz = (f_t(\bfx),\bfx)$, and outputs a vector $f_{c_k}(\bfz)$ that corresponds to the first occurrence of event $k$. Together all these outputs provide the estimates $\hat{P}(t, \delta^* | \bfx)$, which in turn, can be used to estimate the CI, $F_k(t)$ for each event type $k$ as
$$
\widehat{F}_k(t|\bfx) = \sum_{t^*=0}^t \hat{P}(t^*, \delta^* | \bfx).
$$

This estimated CI is used to compare and assess model performance, particularly in evaluating how well the model discriminates among cause-specific risks.

The DH model is trained using a loss function that incorporates both the likelihood of observed survival times and an additional ranking loss to ensure accurate ranking of event times, expressed as $\mathcal{L}_{Total}=\mathcal{L}_1 + \mathcal{L}_2$. The log-likelihood loss $\mathcal{L}_1$ measures the likelihood of the observed survival times, including the censoring information and is given by
\begin{align*}
\mathcal{L}_1 =& -\sum_{i=1}^{n} \left[ 1(\delta_i^* \ne 0)  \log\left(y_{\delta_i^*,t_i}\right) \right. \\
&\left. + 1(\delta_i^* = 0) \log\left( 1 - \sum_{k=1}^K F_k(t_i | \bfx_i) \right) \right],
\end{align*}
The ranking loss $\mathcal{L}_2$ helps the model to correctly rank the event times of pairs of individuals and is given by
\begin{equation}
\begin{split}
\mathcal{L}_2 = \sum_{k=1}^{K} \alpha_k \cdot \sum_{i \neq j} A_{k, i, j} \cdot \eta\left( F_k(t_i | \bfx_i), F_k(t_i | \bfx_j) \right)
\end{split}
\end{equation}
where 
$$
A_{k, i, j} = 1(\delta_i^*=k, T_i<T_j)
$$ 
is an indicator function that identifies acceptable pairs for comparison for event type $k$, $\eta(x,y)$ is a convex loss function, and $\alpha_k$ determines ranking losses for event type $k$. Specifically, \citep{lee_et_al_2018} considers $\alpha_k=\alpha$ for $k=1,2,\ldots,k$ and $\eta(x,y)=\exp(-(x-y)/\sigma)$.

%☀️☀️☀️
\section{Simulation study} \label{sec:simulation}
We conducted an extensive simulation analysis to compare the performance of the considered competing risk analysis methods under diverse data-generating scenario with the objective to identify for different scenarios, the unique strengths and weaknesses of the approaches, and provide a guideline for the practitioners. The following sections describe data generation procedure, performance evaluation metrics, method implementation details, and results.

%☀️☀️
\subsection{Data Generation}
We generated survival data for $n=200$, $300$, $500$, $1000$ subjects, each having two competing risks and values of $p=24$, $212$, $512$ and $1012$ covariates, based on the Fine–Gray (PSDH) model, where event 1 was considered the primary event. The first twelve covariates from $X_1$ to $X_{12}$ were considered as true covariates, including six continuous covariates from $X_1$ to $X_{6}$ and six binary covariates from $X_7$ to $X_{12}$. The true covariates had non-zero coeficients (given below), and the coeficients of the rest of the covariates were set as zero.

The continuous covariates were generated with three different correlation structures:
\begin{enumerate}[topsep=0pt]
\item Independent: $X_j \sim N(0,1)$, $j=1,2,\ldots,6$
\item Exchangeable: $(X_1,\ldots,X_6) \sim N_6(0,V)$, where $V$ was a block-diagonal covariance matrix with two exchangeable blocks of three covariates each, where within each block the diagonal elements were 1 and pairwise correlation was $\rho_{ii'}=r$ with $r=0.2$, $0.5$, $0.8$. 
\item Autoregressive order 1 (AR(1)):  Same structure as exchangeable with the difference of $\rho_{ii'}=r^{|i-i'|}$ with $r=0.2$, $0.5$, $0.8$.
\end{enumerate}

For generating the binary covariates from $X_7$ to $X_{12}$, we generated continuous covariates in the same ways as mentioned above, and then converted them into binary form using the dichotomization:
$X_j =1$ if $X_j < r_b$ and $X_j=0$ otherwise, $j=7,8,\ldots,12$,
with
\begin{enumerate}[topsep=0pt]
\item $r_b=0$, which gives a balanced binary distribution;
\item $r_b=-1$, which gives a binary distribution with sparser $1$s.
\end{enumerate}

We used three models for the linear predictor in the exponent of the Fine–Gray model:
\begin{enumerate}[topsep=0pt]
\item Linear: This model included linear terms in the true covariates, i.e,
$$ 
\beta_1 X_1 + \beta_2 X_2 + \dots + \beta_{12} X_{12},
$$
where \\
$\bfbeta = (\log(2), -\log(2), 0, 0, \log(2), -\log(2), \\1.5, -1.5, 0, 0, 1.5, -1.5)$ for event 1 and \\$\bfbeta = (0, 0, \log(2), -\log(2), \log(2), -\log(2), \\0, 0, 1.5, -1.5, 1.5, -1.5)$ for event 2;

\item Quadratic: This model included linear terms in the true covariates and quadratic terms in the continuous covariates, i.e., 
\begin{align*}
&\beta_1 X_1 + \beta_2 X_2 + \dots + \beta_{12} X_{12} + \beta_1^Q X_1^2 + \beta_2^Q X_2^2 \\
&+ \dots + \beta_6^Q X_6^2,
\end{align*}
where\\
$\bfbeta^Q = (\log(2), -\log(2), 0, 0, \log(2), -\log(2))$ for cause 1 and\\
$\bfbeta^Q = (0, 0, \log(2), -\log(2), \log(2), -\log(2))$ for cause 2;

\item Interaction: This model included linear terms in the true covariates and interaction terms of the form $I(X_k>0)\cdot X_{k+6}$, $k=1,2,\ldots,6$ between the continuous and binary covariates, i.e., 
\begin{align*}
&\beta_1 X_1 + \beta_2 X_2 + \dots + \beta_{12} X_{12} + \beta_1^I I(X_1>0)X_7 \\
&+ \beta_2^I I(X_2>0)X_8 + \dots + \beta_6^I I(X_6>0)X_{12},
\end{align*}
where\\
$\bfbeta^I = (-\log(2), \log(2), 0, 0, -\log(2), \log(2))$ for cause 1 and\\
$\bfbeta^I = (0, 0, -\log(2), \log(2), -\log(2), \log(2))$ for cause 2;
\end{enumerate}

The parameter values in the Fine–Gray model were chosen to reflect realistic and interpretable effect sizes. Specifically, regression coefficients of $\log(2)$ and $-\log(2)$ correspond to moderate hazard ratios of approximately 2 and 0.5, while coefficients of $\pm 1.5$ represent stronger effects, allowing us to evaluate method performance across a range of signal strengths.

The above specifications include 4 sample sizes ($n$), 4 numbers of covariates ($p$), 3 correlation structures for continuous covariates (cortype) and 3 correlation strengths for dependent structures ($r$), 2 sparsity structures for binary covariates $(r_b)$, and 3 predictor models (model). Together, these factors result in 672 unique simulation scenarios. For each scenario, we generated 10 replicates. Event times ($T_i^*$) were simulated from the Fine–Gray model, while censoring times ($C_i$) were independently drawn from a uniform distribution, $C_i \sim \text{Unif}(0, 20)$. The \emph{simulateTwoCauseFineGrayModel} function from the R package \emph{fastcmprsk} \citep{fastcmprsk} was utilized for data generation. In the simulated datasets, the average censoring rate across scenarios was about 25\% (SD $\approx$ 7\%).
%For the exponential censoring, the times was selected from an exponential distribution \(C_i \sim \text{Exp}(0.01)\), but only values $c_i \leq 20$ was kept.

Importantly, our simulation design departs from prior comparative studies involving the methods considered in this work (e.g., \citet{fu_et_al_2017,ishwaran_et_al_2014,binder_et_al_2009}) by spanning a broader grid of sample sizes and dimensionalities, incorporating richer covariate types and correlation structures, and evaluating a more diverse set of performance metrics. Supplementary Table~1 contrasts our design with these studies in terms of simulation settings, models compared, and evaluation metrics.

%☀️☀️
\subsection{Evaluation metric}
We used six metrics to evaluate and compare the performance of the competing risk methods described in Section~\ref{sec:methods}. The metrics are described below
\begin{enumerate}
\item True Positive Rate ($TPR$): The proportion of true covariates that are correctly identified, given by
$$
{TPR} = \frac{TP}{TP + FN},
$$
where $TP$ is the number of true positives and $FN$ is the number of false negatives.

\item False Discovery Rate ($FDR$): The proportion of false positives among all positive predictions, given by
$$
{FDR} = \frac{FP}{FP + TP},
$$
where $FP$ is the number of false positives.

\item Beta Error ($betaerr$): The sum of squared differences between the true coefficients ($\beta_j$s) and the estimated coefficients ($\hat{\beta}_j$s) 
%\citep{zou2005regularization}:
$$
{betaerr} = \sum_{j=i}^J \left( \hat{\beta}_j - \beta_j \right)^2.
$$

\item Concordance Index ($cindex$): The proportion of concordant pairs of subjects  $i$ and $j$, where the subject with the shorter survival time has a higher predicted risk \cite{harrell1996multivariable}, given by
$$
{cindex} = \frac{\sum I\left( \hat{R}_i > \hat{R}_j \right) \cdot I\left( T_i < T_j \right)}{\text{Number of comparable pairs}},
$$
where $\hat{R}_i$ and $\hat{R}_j$ are the predicted risks (cumulative incidence function) for individuals $i$ and $j$.

\item Time-dependent Area Under the Curve ($AUC_t$): $AUC_t(t^*)$ is the probability, at time $t^*$, that, for a randomly selected pair of individuals, the predicted risk score of the individual who experiences the event before time $t$ is higher than the predicted risk score of the individual who experiences the event after time $t$ \cite{blanche_et_al_2019,saha_heagerty_2010}:
$$
AUC_t(t^*) = P\left( \hat{R}_i > \hat{R}_j \,\middle|\, T_i \le t^* < T_j \right).
$$

\item Integrated Brier Score ($IBS_t$): The time-dependent Brier score averaged over a range of times \cite{graf1999assessment}. The time-dependent Brier score at time $t^*$ is given by:
$$
BS_t(t^*) = \frac{1}{n} \sum_{i=1}^{n} \left( N(t^*) - \widehat{\pi}(t^*|\bfx_i) \right)^2 w_i(t^*;\widehat{G}),
$$
where $\widehat{\pi}(t)$ is estimated CI at time $t$ and $w_i(t)$ are the inverse probability of censoring weights (c.f., Section~\ref{sec:PR}). The Integrated Brier Score at time $t^*$ is:
$$
IBS_t(t^*) = \frac{1}{t^*} \int_0^{t^*} BS_t(u) \, du.
$$
\end{enumerate}

Among these measures, $TPR$ and $FDR$ evaluate variable selection performance, $betaerr$ evaluates estimation performance, $cindex$ and $AUC_t$ evaluate discriminative performance (ranking ability), and $IBS_t$ evaluates calibration performance. However, not all measures are available for all methods under consideration. Both PR and CB methods are based on the Fine–Gray model. So, although PR provides shrinkage estimates in a single optimization step whereas CB is an ensemble method where covariate estimates are built up over iterations, the final covariate estimates from them are somewhat comparable. RF method does not provide any covariate estimates but provides variable importance and minimal depth, which can be used for variable selection. DH does not provide any covariate-specific measures. As a result, $TPR$ and $FDR$ are available only for PR, CB, and RF and $betaerr$ is available only for PR and CB, whereas $cindex$, $AUC_t$, and $IBS_t$ are available for all the methods.

%☀️☀️
\subsection{Method Implementation Details}
The $PR$ methods were implemented using the \emph{crrp} function from the R package \emph{crrp} \citep{crrp}. The optimal penalty was selected using the Bayesian Information Criterion ($BIC$) criterion. The CB method was implemented using \emph{CoxBoost} function from the R package \emph{CoxBoost} \citep{coxboost}, using the \emph{pscore} criterion for penalization. The optimal number of boosting steps was determined using a 10-fold cross-validation. The RF method was implemented using the \emph{rfsrc} function from the R package \emph{randomForestSRC} \citep{rfsrc}, with $100$ trees and splitting based on the Gray's rule (option \emph{logrankCR}). The RF tuning parameters---number of candidate variables randomly selected at each split of a tree and the minimum terminal node size---were optimally determined through a grid-based search using out-of-bag error. The DH model was implemented using Python code built on \emph{TensorFlow/Keras} modules, as provided by the authors \citep{deephit}. The data were randomly split into a training (80\%) and test data, DH model was fit to the training data using 128 nodes and 2 fully connected layers for the shared subnetwork, 64 nodes and 1 fully connected layer for the cause-specific subnetworks, and RELU activation function following \citep{deephit}. The loss hyperparameters---weight between likelihood loss and ranking loss (alpha) and kernel width in ranking loss (sigma)---were tuned using a validation set time-dependent concordance index, with 20\% of training held out for validation. Performance evaluation criteria were computed in the test data.

For PR, CB, and RF methods, The $cindex$ measure was computed using the \emph{cindex} function from the R package \emph{pec} \citep{pec}, whereas $AUC_t$ and $IBS_t$ were computed using the \emph{Score} function from the R package \emph{riskRegression} \citep{riskregression}, all at time $t=10$. For DH, the Python module \emph{sksurv} was used to compute $AUC_t$ and $IBS_t$. 

%☀️☀️
\subsection{Result}
In this section, we present the results of our analysis. As noted in \cite{fu_et_al_2017}, PR methods are designed primarily for $n>p$ scenarios; when applied to $n<p$ data, they may identify true signals but at the cost of excessive false positives and instability. Consistent with this, we could not obtain results for most simulated datasets with $n<p$ using PR. So, for datasets with $n<p$, results only from the other methods will be used. For RF, following \citep{ishwaran_et_al_2014}, covariates that had a positive variable importance value and satisfied a minimal depth threshold determined from the forest were selected. Below, we summarize the results in terms of the various evaluation metrics. 
%As the number of covariates $p$ increases, especially when $p>n$, \textit{LASSO}, \textit{SCAD}, and \textit{MCP} become computationally intensive and took an excessive amount of time, making runtime infeasible to report.

%☀️
\subsubsection{Comparison of variable selection performance}
\begin{figure}
    \includegraphics[width=9cm]{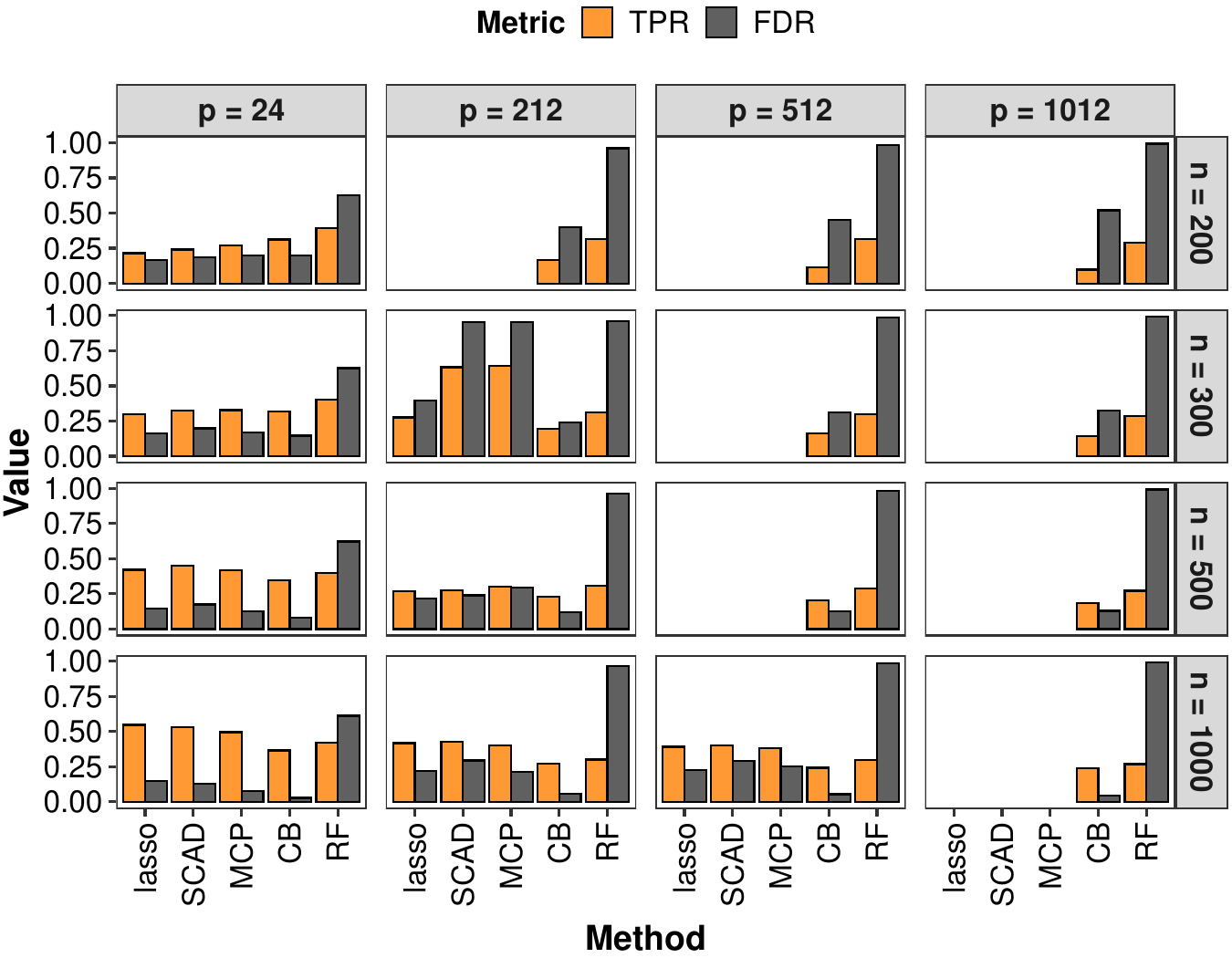}
    \caption{Comparison of variable selection performance: $TPR$ and $FDR$ values across methods (where available) for varying sample size ($n$) and number of covariates ($p$).}
    \label{fig:varsel1}
\end{figure}

\begin{figure}
    \includegraphics[width=9cm]{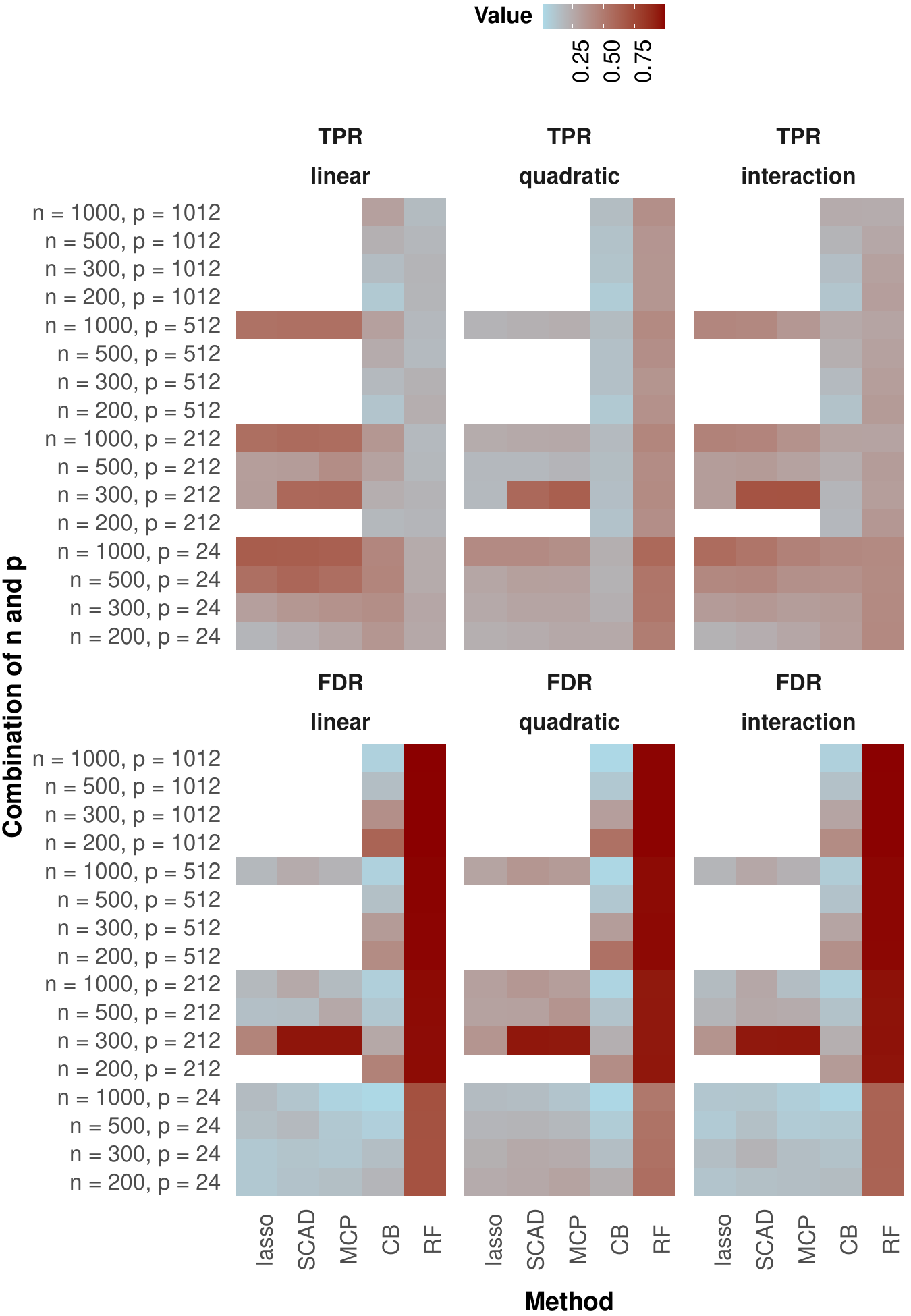}
    \caption{Comparison of variable selection performance: $TPR$ and $FDR$ values across models and methods (where available) for varying sample size ($n$) and number of covariates ($p$).}
    \label{fig:varsel2}
\end{figure}

\begin{figure}
    \includegraphics[width=9cm]{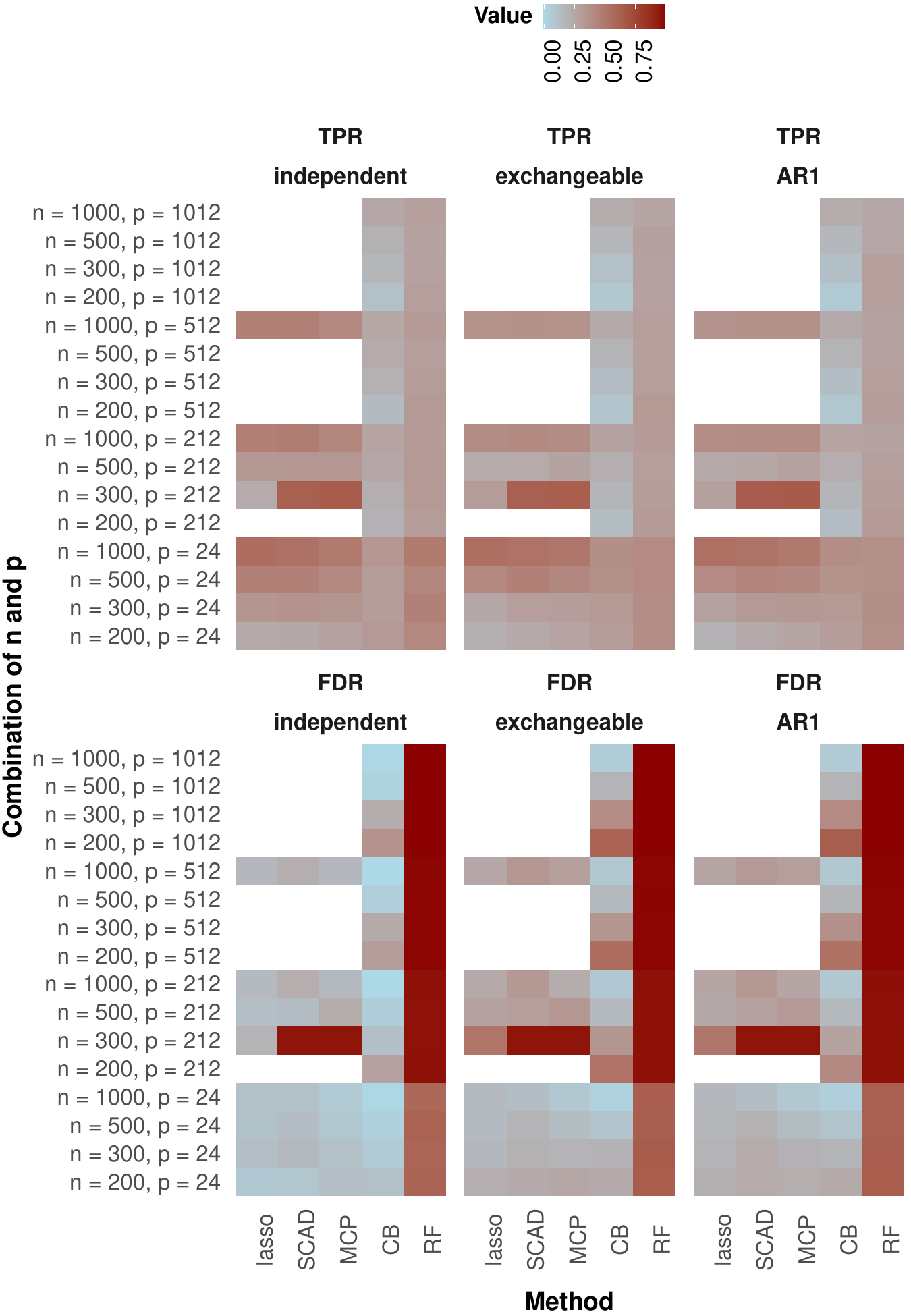}
    \caption{Comparison of variable selection performance: $TPR$ and $FDR$ values across correlation structures (of the continuous covariates) and methods (where available) for varying sample size ($n$) and number of covariates ($p$).}
    \label{fig:varsel3}
\end{figure}

We present the $TPR$ and $FDR$ results in Figure~\ref{fig:varsel1} through Figure~\ref{fig:varsel3}. Specifically, Figure~\ref{fig:varsel1} shows the $TPR$ and $FDR$ values across methods (where available) for varying sample size ($n$) and number of covariates ($p$), averaged over all other specifications and replicates, i.e., over the $(n,p,\text{method})$ grid. Figure~\ref{fig:varsel2} and Figure~\ref{fig:varsel3} present similar plots over finer grids $(n,p,\text{model},\text{method})$ and $(n,p,\text{cortype},\text{method})$, respectively. A higher TPR is better whereas a lower FDR is better. We note that these metrics are most naturally defined for methods that produce explicit sparse coefficient estimates (e.g., PR and CB approaches). For methods such as RF, which rely on importance measures rather than coefficient sparsity, these metrics are not directly comparable and should be interpreted with caution; DH does not yield an explicit variable selection mechanism and is therefore excluded.

\begin{itemize}[topsep=0pt]
\item Figure~\ref{fig:varsel1} clear performance trade-offs across $n$ and $p$. For all penalized regression methods, TPR generally increases with $n$ when $p$ is small, while performance deteriorates as $p$ grows. SCAD and MCP achieve the highest TPR in some moderate-dimensional settings (notably $n=300,p=212$), but this gain is accompanied by extremely high FDR, indicating substantial over-selection. LASSO exhibits more stable behavior, with moderate TPR and lower FDR, though its TPR declines in higher-dimensional regimes. In contrast, the CB method consistently favors sparsity, resulting in lower TPR across all configurations, particularly for large $p$, but substantially improved FDR control as $n$ increases, even when $p\gg n$. Random Forest attains relatively stable but modest TPR across settings, while suffering from uniformly high FDR, especially in high-dimensional scenarios, leading to poor overall variable selection performance. However, these results should be interpreted cautiously, given the importance-based nature of RF selection.

\item Figure~\ref{fig:varsel2} shows that, among the CB variants, the interaction model generally achieves the highest true positive rates across $n$ and $p$, followed closely by the linear model, while the quadratic model consistently underperforms in terms of sensitivity. However, the linear CB model tends to exhibit slightly lower false discovery rates than the interaction model, yielding a more favorable overall balance between detection and false positive control. This suggests that CB is most effective for capturing direct effects and pairwise interactions, with limited benefit from quadratic expansions. In contrast, for Random Forest, the quadratic model consistently outperforms the linear and interaction variants in terms of signal recovery, although RF methods overall remain less competitive due to substantially higher false discovery rates, with the above caveat on comparability.

\item Figure~\ref{fig:varsel3} indicates that CB achieves its best performance under independent correlation structures (CB-Independent), with progressively weaker performance under exchangeable (CB-Exchangeable) and AR(1) (CB-AR(1)) settings. While TPR remains relatively stable across dependence structures, increasing local correlation leads to modestly higher FDR, suggesting that correlated designs introduce additional noise that slightly degrades false discovery control.
\end{itemize}

%Regarding $FDR$, LASSO shows low values for small $p$ but increases with higher $p$, indicating a higher false positive rate, whereas SCAD, MCP, and CB maintain a better balance between $TPR$ and $FDR$. CB emerges as the most efficient method, maximizing $TPR$ while keeping $FDR$ low, while LASSO and RF present trade-offs that may be less suitable for high-dimensional variable selection. 

%☀️
\subsubsection{Comparison of estimation performance}
\begin{figure}
    \includegraphics[width=8cm]{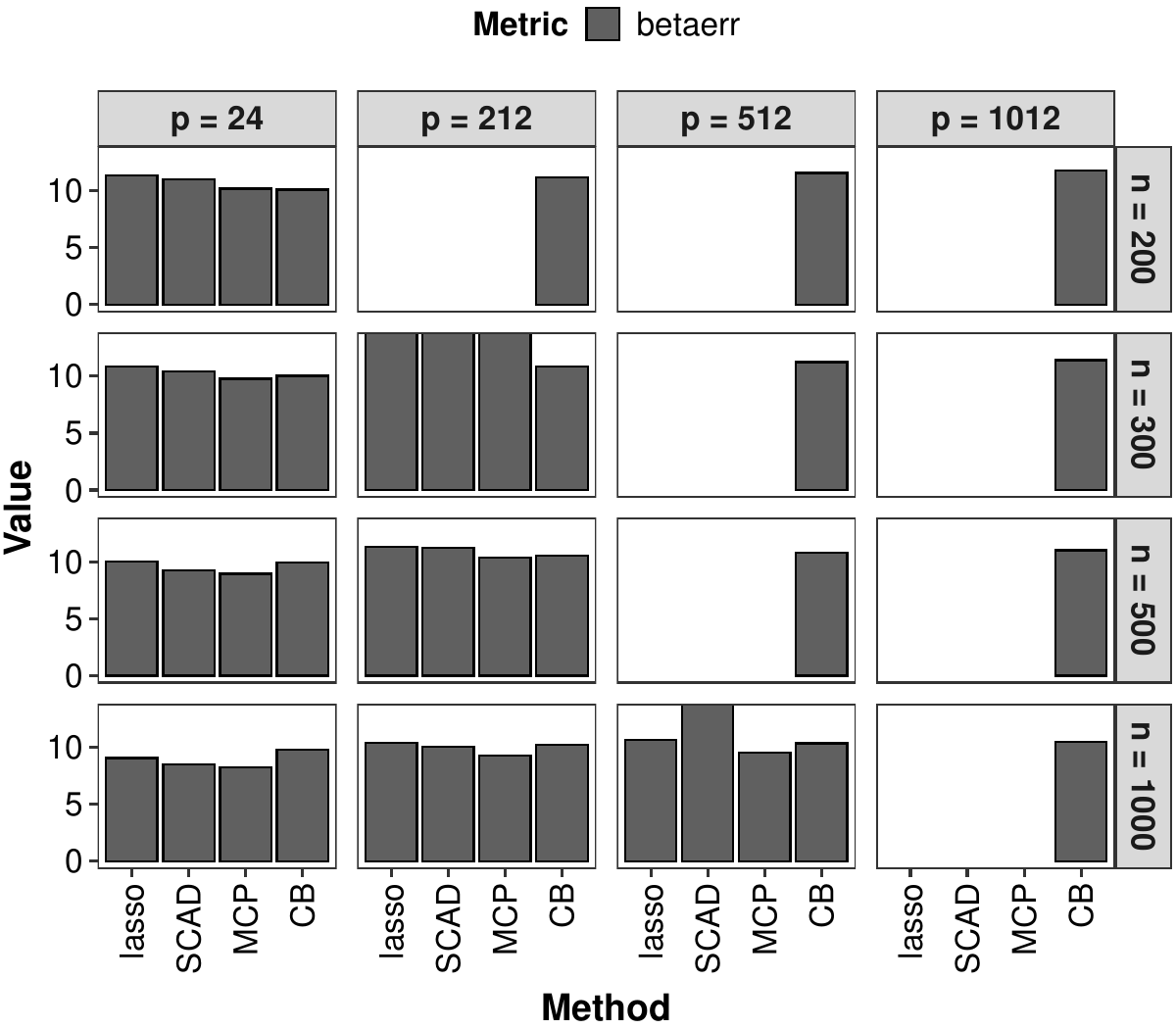}
    \caption{Comparison of estimation performance: $betaerr$ across methods (where available) for varying sample size ($n$) and number of covariates ($p$).}
    \label{fig:estimation}
\end{figure}

Figure~\ref{fig:estimation} compares the estimation error (\emph{betaerr}) of PR and CB methods across varying $n$ and $p$, averaged over remaining specifications. While SCAD and MCP occasionally achieve slightly lower \emph{betaerr} than CB in low-dimensional settings, their estimation error becomes severely inflated for moderate to high $p$, with extreme values not fully displayed in the figure to preserve scale. In contrast, CB exhibits stable and well-controlled \emph{betaerr} across all $n$ and $p$, including high-dimensional regimes. Supplementary Figure~1 and Supplementary Figure~2 further illustrate these patterns across correlation structures and model types. Overall, CB demonstrates more consistent and reliable estimation accuracy than the PR methods, particularly as dimensionality increases.

%☀️
\subsubsection{Comparison of discriminative performance}
\begin{figure}
    \includegraphics[width=9cm]{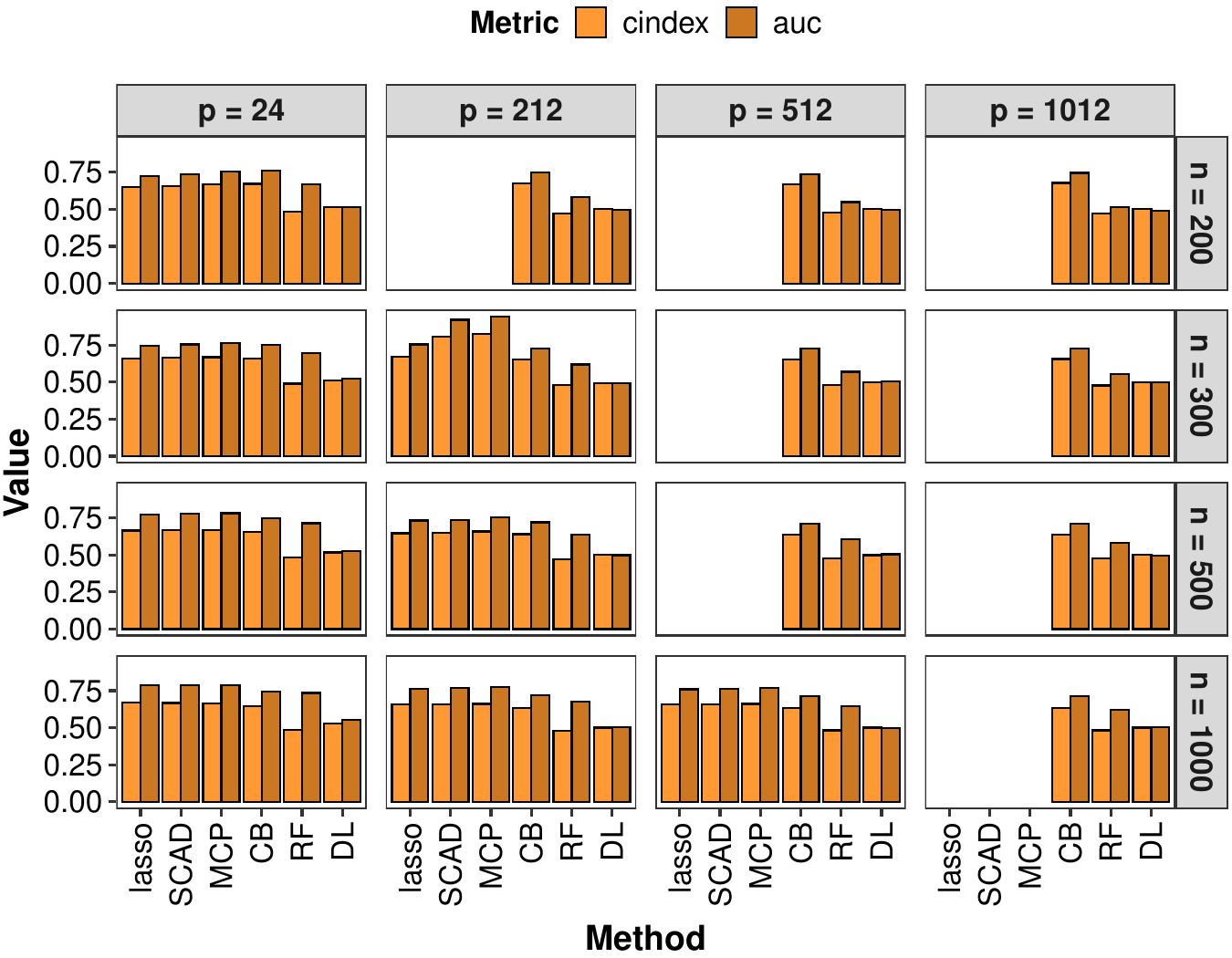}
    \caption{Comparison of discriminative performance: $cindex$ and $AUC$ values across methods (where available) for varying sample size ($n$) and number of covariates ($p$).}
    \label{fig:discrim1}
\end{figure}

\begin{figure}
    \includegraphics[width=9cm]{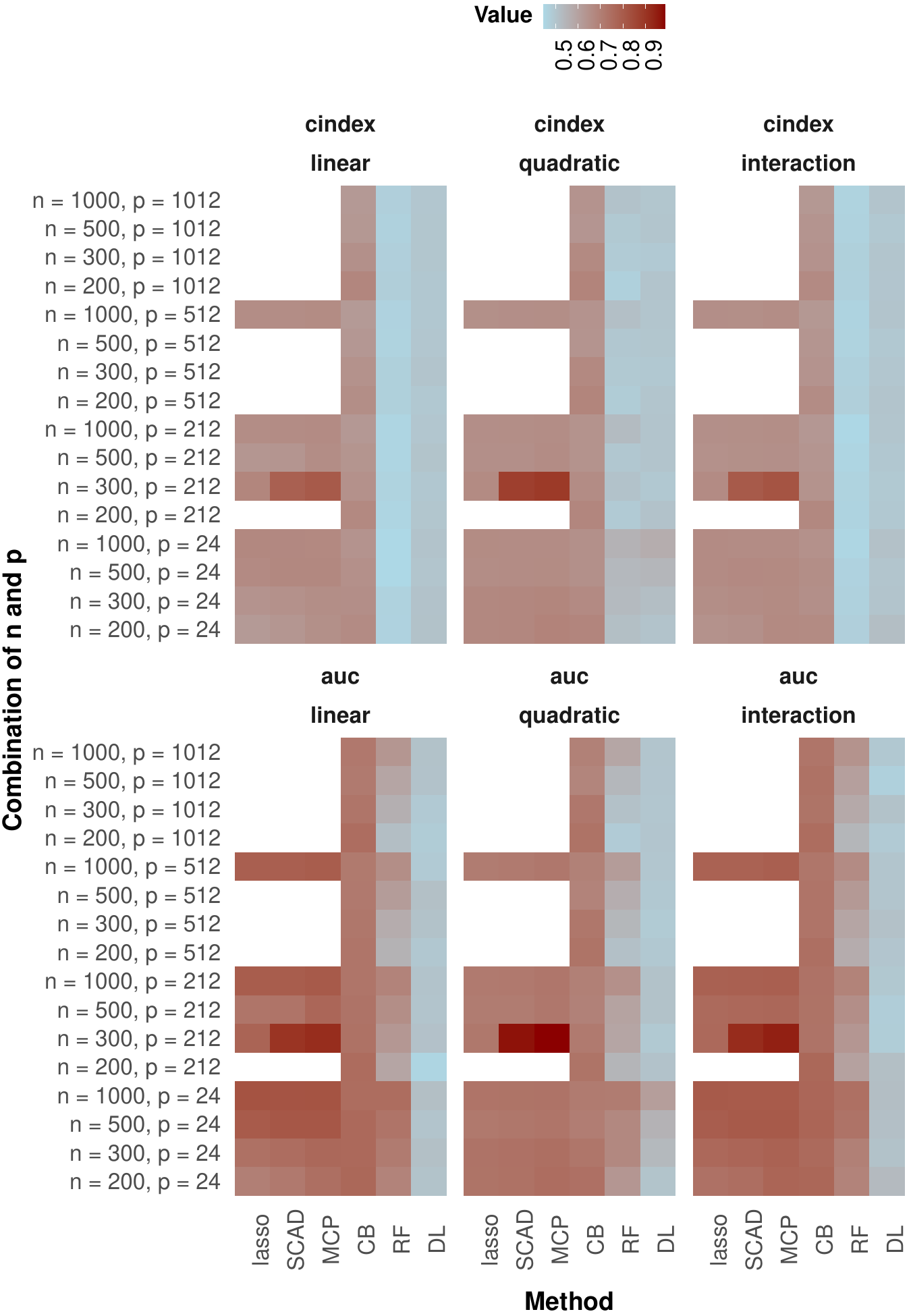}
    \caption{Comparison of discriminative performance: $cindex$ and $AUC$ values across models and methods (where available) for varying sample size ($n$) and number of covariates ($p$).}
    \label{fig:discrim2}
\end{figure}

\begin{figure}
    \includegraphics[width=9cm]{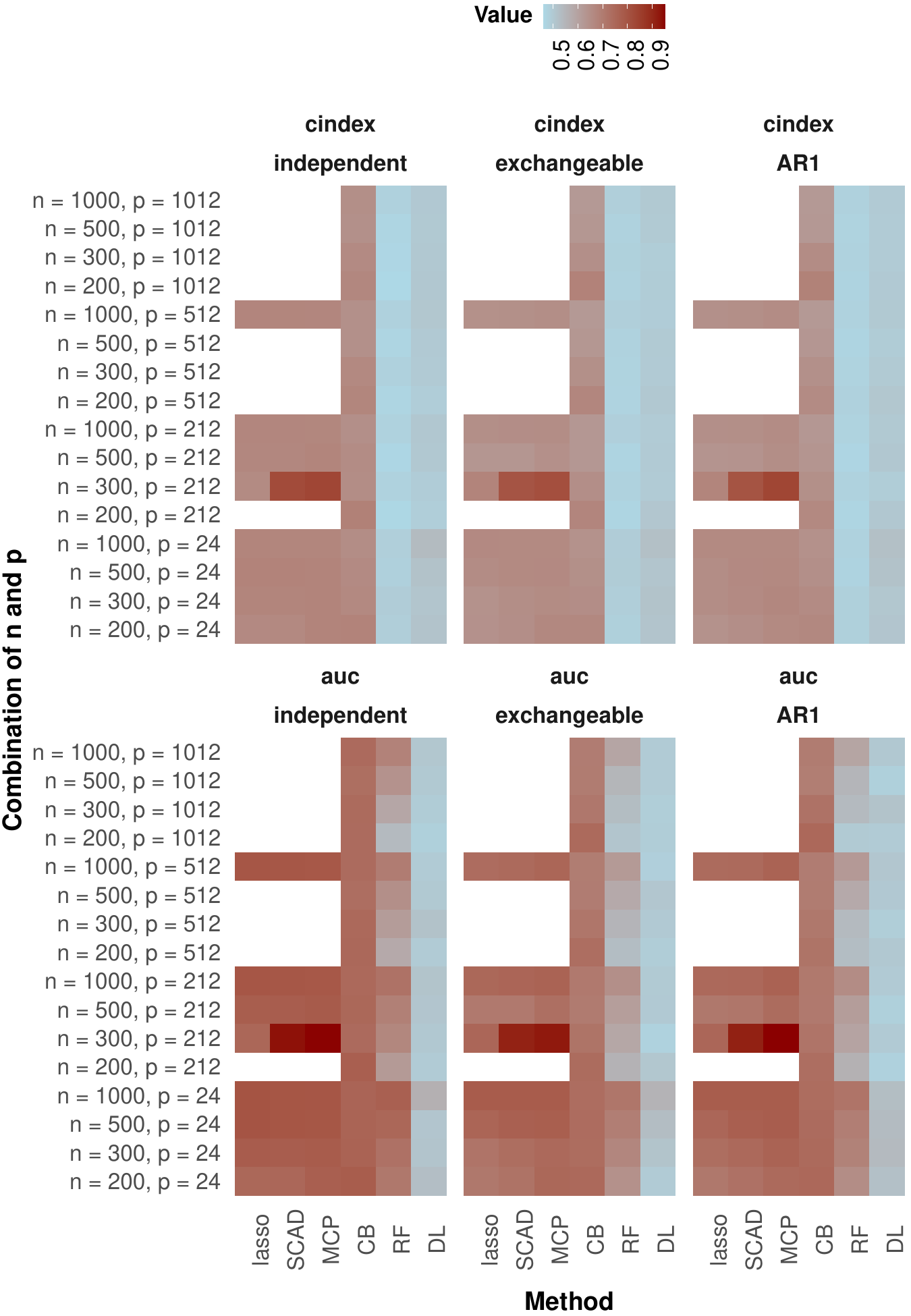}
    \caption{Comparison of discriminative performance: $cindex$ and $AUC$ values across correlation structures (of the continuous covariates) and methods (where available) for varying sample size ($n$) and number of covariates ($p$).}
    \label{fig:discrim3}
\end{figure}

As measures of discriminative performance, $cindex$ and $AUC$ results are presented in Figure~\ref{fig:discrim1} through Figure~\ref{fig:discrim3}. Specifically, Figure~\ref{fig:discrim1} shows the $cindex$ and $AUC$ values over the $(n,p,\text{method})$ grid (where available), averaged over all other specifications and replicates. Figure~\ref{fig:discrim2} and Figure~\ref{fig:discrim3} present similar plots over finer grids $(n,p,\text{model},\text{method})$ and $(n,p,\text{cortype},\text{method})$, respectively. Higher values of $cindex$ and $AUC$ illustrate better discriminative ability.

\begin{itemize}[topsep=0pt]

\item Figure~\ref{fig:discrim1} demonstrates that CB consistently achieves the highest discriminative performance in terms of both \emph{cindex} and \emph{AUC} across all $n$ and $p$, with particularly strong gains in high-dimensional settings. PR methods provide competitive but generally lower discrimination, with occasional improvements in moderate dimensions, while Random Forest exhibits substantially weaker performance, with \emph{cindex} values close to random and notably lower AUC across all scenarios.

\item Figure~\ref{fig:discrim2} shows that for CB methods, the linear and interaction models generally outperform the quadratic model in terms of both \emph{cindex} and \emph{AUC}, particularly as dimensionality increases. A similar pattern is observed for penalized regression methods, where quadratic expansions tend to offer no improvement and often degrade discriminative performance. In contrast, Random Forest exhibits comparatively better discrimination under quadratic specifications, although its overall performance remains inferior to CB and PR approaches.

\item Figure~\ref{fig:discrim3} shows that CB exhibits only modest variation in discriminative performance across independent, exchangeable, and AR(1) correlation structures, with the independent setting yielding slightly higher 
\emph{cindex} and \emph{AUC} in most configurations. Overall, the differences across correlation types are small, indicating that CB maintains robust discriminative ability even in the presence of correlated covariates.

\end{itemize}

%☀️
\subsubsection{Comparison of calibration performance}
\begin{figure}
    \includegraphics[width=9cm]{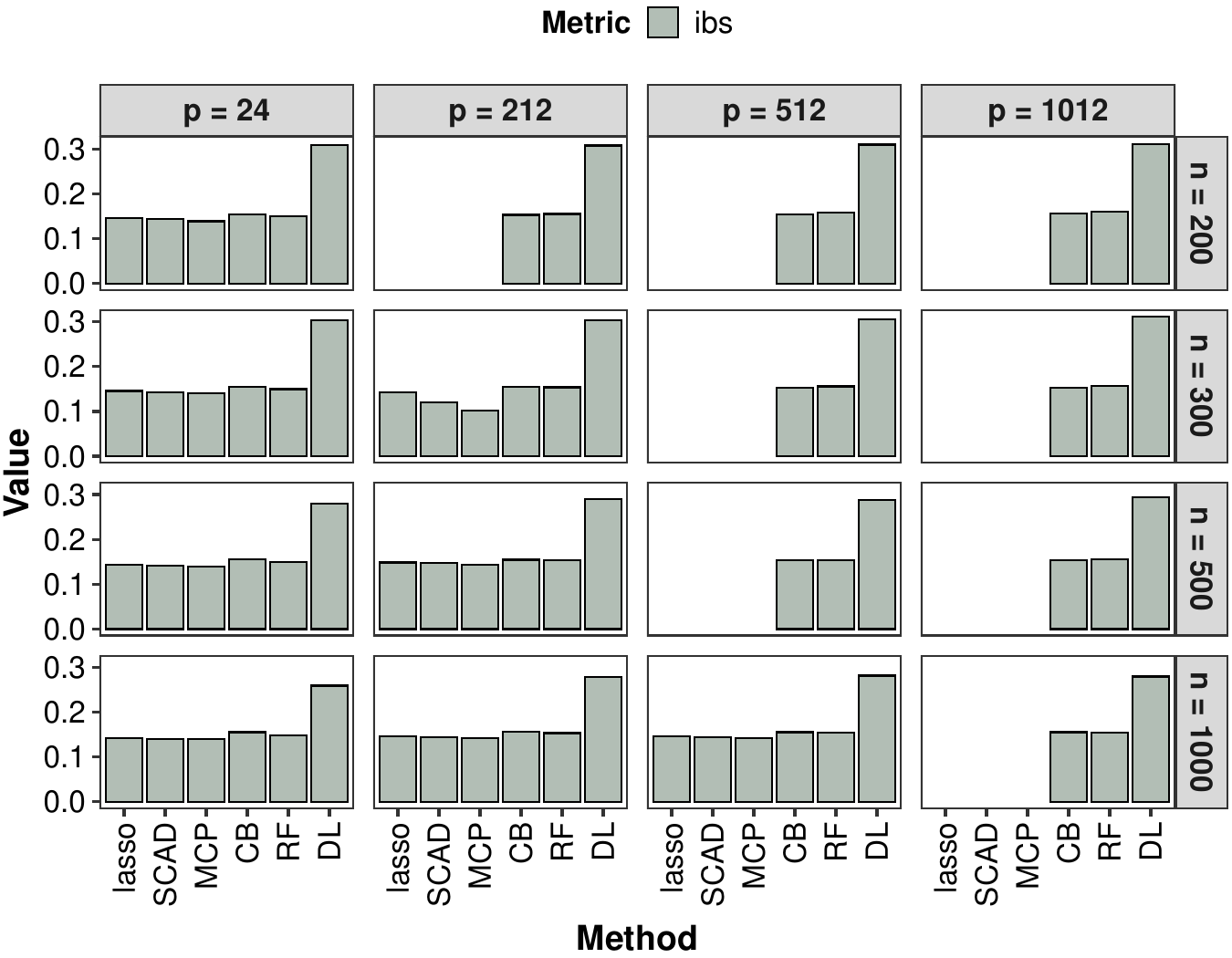}
    \caption{Comparison of calibration performance: $IBS$ across methods (where available) for varying sample size ($n$) and number of covariates ($p$).}
    \label{fig:calibration}
\end{figure}

We assess calibration performance based on $IBS$ where a lower value indicates superior performance. Figure~\ref{fig:calibration} presents IBS values averaged over all other settings for each $(n,p)$ combination. Among the penalized regression methods, MCP consistently attains the lowest \emph{IBS} across nearly all settings, followed by SCAD and then LASSO. Overall, PR methods exhibit superior calibration performance relative to CB and RF, which show comparable but uniformly higher \emph{IBS} values. RF performance further degrades as dimensionality increases, while CB remains stable but less well calibrated. These results indicate that DH exhibits the poorest calibration among all methods considered. We provide $IBS$ plots over $(n,p,\text{cortype},\text{method})$ and $(n,p,\text{model},\text{method})$ in Supplementary Figure~3 and Supplementary Figure~4, respectively.

Additional numerical results, including summary statistics across replicates (reported as mean $\pm$ standard deviation) and sensitivity analyses by correlation structure and covariate model, are provided in Supplementary Tables 2–4.

%☀️
\subsubsection{Overall comparison and guidelines}
We have confirmed that, generally for all settings, both increased strength of correlation among the covariates (determined by $\rho_{ii'}$) and increased sparsity of the binary covariates (determined by $r_{b}$) lead to worse performance for all the methods, which justifies averaging results over $\rho_{ii'}$ and $r_{b}$ in all the plots.

These conclusions should be interpreted in light of the specific implementations and tuning strategies employed in this study, rather than as definitive statements about the broader methodological classes.

The comparative evaluation across variable selection, estimation accuracy, discriminative ability, and calibration highlights clear method-specific trade-offs. CB consistently provides strong false discovery control, stable estimation, and competitive discriminative ability across a wide range of $(n,p)$ configurations, particularly in high-dimensional settings. PR methods (LASSO, SCAD, and MCP) can achieve higher sensitivity or better calibration in selected low- to moderate-dimensional regimes, but may suffer from inflated false discoveries or unstable estimation as dimensionality increases. RF and DH generally underperform relative to CB and PR methods across most metrics in the present implementation, with RF showing weak variable selection and discrimination and DH exhibiting poor calibration; however, these results may depend on tuning choices, particularly for DH. Based on these findings, we summarize the following practical guidelines:

\begin{itemize}
\item If the primary goal is reliable coefficient-based variable selection, stable estimation, and strong discriminative performance, particularly in moderate- to high-dimensional settings ($p \gtrsim n$), CB is recommended.

\item If accurate probability calibration is the primary concern and dimensionality is moderate ($n > p$), PR methods—particularly MCP, followed by SCAD and LASSO—may be preferred; however, their performance may deteriorate as $p$ increases.

\item If the underlying relationships are expected to be highly nonlinear or involve complex interactions, RF and DH may offer advantages; however, in high-dimensional settings with limited sample sizes, their performance may be less stable, and in the present implementation, they show weaker variable selection, discrimination, and calibration performance.
\end{itemize}

Overall, CB offers a robust and well-balanced performance across metrics under the considered settings, while PR methods may be preferred in applications where calibration dominates and dimensionality remains limited. In practice, the choice of method should be guided by the primary analytical objective, the dimensionality of the data, and the anticipated complexity of the underlying relationships.

To assess computational feasibility, we recorded average runtimes across the (n,p) grid. PR runtimes with cross-validation based tuning (for datasets with $n>p$) were modest in low-dimensional settings, remaining under 5 minutes when $p\le 212$, but increased sharply with larger $p$, reaching 80-100 minutes for $n=1000,p=512$. In contrast, CoxBoost with tuning exhibited more favorable scaling with $p$, with runtimes increasing steadily but remaining within 15 minutes (2 minutes without tuning) even for the largest setting $n=1000,p=1012$. In contrast, RF runtimes with tuning increased substantially with both $n$ and $p$, exceeding 100 minutes for $p=1012$ even at $n=200$ ($\approx$106 minutes) and reaching over 200 minutes at $n=1000, p=1012$ ($\approx$209 minutes) in the most demanding scenarios, although, without tuning, RF would require moderate runtimes (approximately 0.1-2.5 minutes. DH with tuning was consistently the fastest ($<0.15$ minutes). Overall, although PR methods became costly in very high-dimensional settings, all approaches were computationally feasible in the scenarios where they were applied. To illustrate this, in Supplementary Figure 5, we provided a visual comparison of runtimes across methods implemented in R over the $(n,p)$ grid, excluding DH, which was implemented in Python and was uniformly faster, but with uniformly non-competitive results.

%☀️☀️☀️
\section{Melanoma Data Analysis} \label{sec:realdata}
In this section, we analyze a Melanoma data using the CB, RF, and DH methods and discuss the results. Detailed information about this dataset is available in \citep{cirenajwis_et_al_2015}. This dataset includes gene-expression profiles obtained from tumor tissue samples collected retrospectively from 214 Cutaneous malignant melanoma (CMM) patients at a single clinical institution. Additionally, it contains survival outcomes and several other covariates for most patients. 

For this analysis, death due to melanoma was considered the primary event of interest, while death from other causes was treated as a competing event. Patients for whom survival data for both events was unavailable were excluded from the analysis. Those who did not die from either events were considered censored. The covariates included in the competing risk model consisted of gene expression levels for $47,323$ genes, along with gender, age, tumor stage, and tissue type. Table~\ref{tab:patient_charecteristics} provides a summary of patient characteristics, including survival status and the covariates gender, age, tumor stage, and tissue type.

\setlength{\tabcolsep}{3pt} % Default is usually 6pt
\begin{table*}[h]
    \renewcommand{\arraystretch}{1.2}
    \caption{Melanoma data: Summary of patient characteristics including survival status, gender, age, tumor stage, and tissue type.}
    \resizebox{\textwidth}{!}{
    \begin{tabular}{l c c c c c c c c c c c c c c}
        \toprule
        \textbf{Patients} & \textbf{Counts} & \multicolumn{2}{c}{\textbf{Gender (\%)}} & \textbf{Age} & \multicolumn{5}{c}{\textbf{Tumor stage (\%)}} & \multicolumn{5}{c}{\textbf{Tissue (\%)}} \\
        \cmidrule(lr){3-4} \cmidrule(lr){6-10} \cmidrule(lr){11-15}
        & & Male & Female & (average) & General & In-transit & Local & Primary & Regional & Cutaneous & Lymph node & Other & Subcutaneous & Visceral \\
        \midrule
        Censored & 60 & 43.3 & 56.7 & 62.9 & 0 & 5 & 10 & 18.3 & 66.7 & 23.3 & 66.7 & 0 & 10 & 0 \\
        Death from Melanoma & 92 & 68.5 & 31.5 & 60.6 & 16.3 & 10.9 & 4.3 & 1.1 & 67.4 & 5 & 103.3 & 1.7 & 33.3 & 10 \\
        Death from other causes & 24 & 62.5 & 37.5 & 66.4 & 0 & 8.3 & 4.2 & 12.5 & 75 & 5 & 28.3 & 0 & 6.7 & 0 \\
        All patients & 176 & 59.1 & 40.9 & 62.1 & 8.5 & 8.5 & 6.3 & 8.5 & 68.2 & 11.4 & 67.6 & 0.6 & 17 & 3.4 \\
        \bottomrule
    \end{tabular}
    }
    \label{tab:patient_charecteristics}
\end{table*}

We analyzed the melanoma data using the CB, RF, and DH methods. For the CB method, we determined the optimal number of boosting steps through a 10-fold cross-validation and used a linear scheme for changing step sizes, where all covariates underwent penalized selection in each step. For RF, we determined the optimal \emph{nodesize} using out-of-bag error via a guarded golden-section line search with noise control, modified Gray splitting rule with uniform weights across time, and random left/right assignments at splits. For DH, the same loss hyperparameter tuning procedure and network architecture were followed as detailed in the simulation analysis.

\begin{figure}
	\includegraphics[width=9cm]{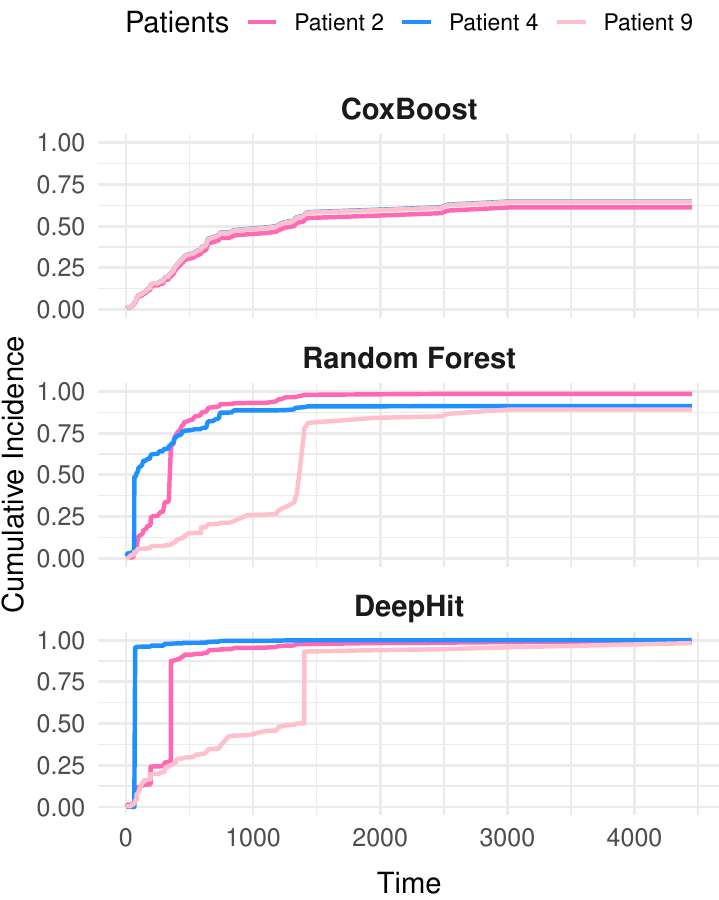}
	\caption{Cumulative incidence estimates for the first three patients in the dataset who succumbed to melanoma at all unique death time points using CB, RF, and DH analyses.}
    \label{fig:CI-melanoma}
\end{figure}

Figure~\ref{fig:CI-melanoma} presents the CI estimates for the first three patients in the dataset who succumbed to melanoma---Patient 2 (a 39-year-old female diagnosed with regional melanoma based on lymph node tissue analysis), Patient 4 (46-year-old male diagnosed with general melanoma based on visceral tissue analysis), and Patient 9 (77-year-old female diagnosed with regional melanoma based on lymph node tissue analysis). The CI estimates are plotted at all unique death time points using CB, RF, and DH method-based analyses.  Across all methods, the plots exhibit a sharp initial increase followed by a plateau, indicating that early event occurrences taper off over time. Notably, the separation among subjects is most pronounced in the RF and DH models, whereas CB maintains a more consistent trend across patients. Overall, CB demonstrates a steady, well-calibrated CI estimate, suggesting it may be the most reliable approach.

\begin{figure}
	\includegraphics[width=9cm]{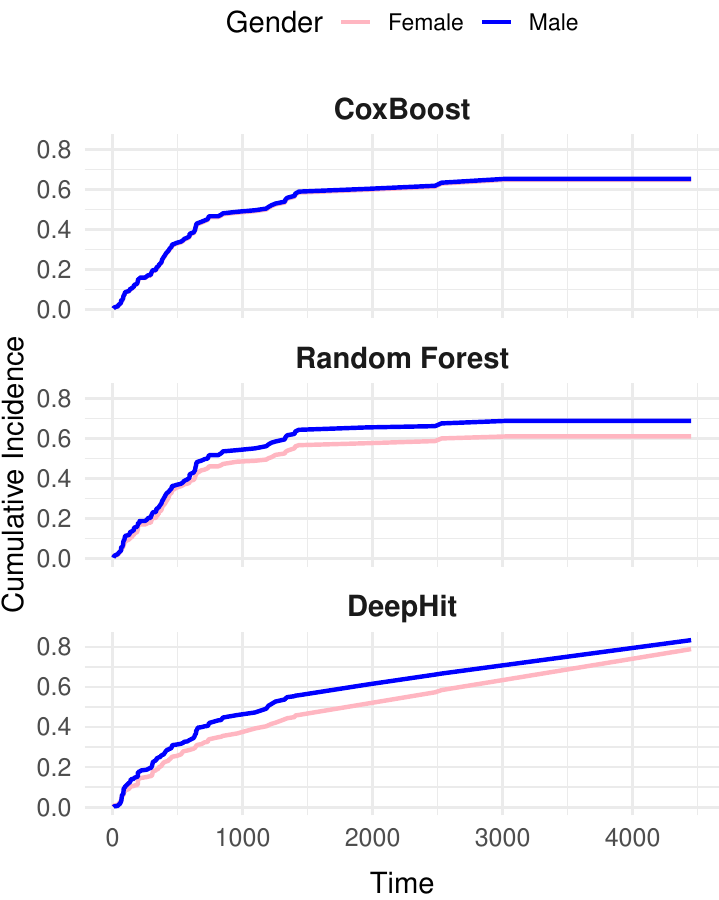}
	\caption{Cumulative incidence estimates averaged over males and females who succumbed to melanoma at all unique death time points using CB, RF, and DH analyses.}
    \label{fig:CI-melanoma2}
\end{figure}

In Figure~\ref{fig:CI-melanoma2}, we present the CI estimates averaged over male and female patients who succumbed to melanoma. The degree of separation between genders varies across the three models, with CB showing no significant difference, while RF and DH indicate that females have a slightly lower risk of melanoma.

CB selected 1 covariate (the one with non-zero estimated coefficient), a gene expression corresponding to the probe ID ILMN\_1727023. RF selected 1266 covariates (the ones that had positive variable importance and satisfied a data-driven minimal depth threshold), all of which were gene expressions, but not including the CB-selected one. %These two sets of selected covariates contained 1 common covariate, , which appears among the first 50 RF selected covariates. 
Table~\ref{tab:genes} shows the CB-selected gene expression and top 10 RF-selected gene expressions, sorted by their variable importance measure, along with the names of the genes they map to (when available) based on the Illumina HumanHT-12 v4.0 annotations.

\begin{table}
\caption{Covariates (gene expressions) selected using CB and RF (top 10, in order of importance) analysis of the Melanoma data} \label{tab:genes}
\begin{tabular}{p{3.5cm}l p{2.5cm}}
Probe ID & Gene name \\
\hline
CB: & & \\
ILMN\_1727023 & AMMECR1L \\
RF: & & \\
ILMN\_1695378 & SREBF1 \\
ILMN\_1786045 & SLC24A5 \\
ILMN\_2054928 & XPNPEP3 \\
ILMN\_1772081 & RPL23AP82 \\
ILMN\_1719280 & C9orf163 \\
ILMN\_1676924 & CD247 \\
ILMN\_2259155 & OTOL1 \\
ILMN\_1738276 & TMEM185A \\
ILMN\_1878644 & -- \\
ILMN\_1677511 & PTGS2 \\
\end{tabular}
\end{table}

Among these genes, \emph{PTGS2}, \emph{SREBF1}, \emph{SLC24A5}, \emph{XPNPEP3}, and \emph{CD247} have been associated with melanoma. \emph{PTGS2} (COX-2) is one of the most consistently implicated inflammatory genes in melanoma. Multiple studies report elevated \emph{PTGS2} expression in primary and metastatic melanoma compared with normal melanocytes, linking it to tumor progression, invasion, and poor prognosis \citep{goulet_et_al_2003, zelenay_et_al_2015, panza_et_al_2016, jessen_et_al_2020}. While COX-2 is largely absent in benign nevi and primary melanoma, it is strongly upregulated in metastatic disease, particularly in lymph node metastases and melanoma cell lines \citep{goulet_et_al_2003}. High COX-2 expression in metastatic melanoma is associated with markedly reduced progression-free survival, independent of BRAF or NRAS mutation status \citep{panza_et_al_2016}. Mechanistically, COX-2 promotes melanoma growth and invasiveness via prostaglandin E2–mediated immune suppression and tumor-promoting inflammation \citep{zelenay_et_al_2015}, with NRF2-driven COX-2 induction linking oxidative stress to immune evasion and increased malignancy \citep{jessen_et_al_2020}. \emph{SREBF1} has been reported to contribute to elevated de novo fatty-acid synthesis programs in melanoma cells \citep{wu_anders_2019}. \emph{SLC24A5} has been implicated in melanoma risk through pigmentation-related genetic variation, with the well-characterized rs1426654 variant showing significant associations in melanoma susceptibility analyses \citep{reis_et_al_2020}. \emph{XPNPEP3} has been implicated in melanoma through functional screening and in vivo metastasis studies, including work identifying its involvement in melanoma-secreted amyloid-$\beta$–mediated brain metastasis \citep{kleffman_et_al_2022}. \emph{CD247} has been characterized as a key T-cell signaling marker in melanoma and has been reported as a biomarker associated with immunotherapy response and clinical prognosis in melanoma patients \citep{zhang_et_al_2023, vo_et_al_2022}. Although we did not find any clear melanoma-focused paper for \emph{AMMECR1L}, it appears in melanoma resources such as the Human Protein Atlas. We could not find any direct and clear associations of the genes \emph{RPL23AP82}, \emph{C9orf163}, \emph{OTOL1}, and \emph{TMEM185A} with melanoma.

Taken together, these findings illustrate a contrast between methods: CB yielded a sparse set of covariates, while RF produced a much larger set, including metastasis- and pigmentation-related candidates such as \emph{PTGS2}, \emph{SLC24A5} and \emph{XPNPEP3}. This suggests that CB may favor parsimony and interpretability by focusing on a few key predictors, whereas RF may prioritize comprehensive signal detection, potentially capturing additional candidates at the expense of parsimony.

However, we note that the melanoma data analysis is primarily presented as an illustrative example to demonstrate the application of the methods in a high-dimensional real-world setting, rather than as a source of new biological discoveries. Although almost half of the genes presented in Table~\ref{tab:genes} have documented biological relevance to melanoma, overall, the findings were not biologically validated in this study. Therefore, this analysis should be interpreted strictly as a methodological use-case demonstration rather than evidence of practical or clinical superiority; future work will be needed to incorporate resampling-based validation, independent datasets, and clinically meaningful utility assessments.

%☀️☀️☀️
\section{Discussion and Conclusion} \label{sec:conclusion}
In this work, we provide a comprehensive review and comparison of some modern competing risk analysis methods in both low- and high-dimensional settings, addressing a key gap in the literature. Our study draws strength from a diverse range of data-generating conditions and the use of multiple evaluation metrics, assessing variable selection, estimation accuracy, discrimination, and calibration performance. The findings highlight the strengths and weaknesses of different approaches, offering valuable insights for researchers and practitioners working with high-dimensional competing risk data. 

In summary, our results show that CB consistently delivers strong false discovery control, stable estimation, and competitive discriminative performance, particularly in high-dimensional scenarios, under the considered settings. PR methods (LASSO, SCAD, MCP) can achieve improved calibration or sensitivity in low- to moderate-dimensional regimes with $n>p$ but may exhibit increased variability or inflated false discoveries as dimensionality increases. RF and DH are effective at capturing nonlinear effects and complex interactions; however, in the settings considered here and under the specific implementations used, they tend to show weaker variable selection and calibration relative to CB and PR. In particular, variable selection comparisons involving RF should be interpreted with caution due to its reliance on importance measures rather than coefficient sparsity. Their performance is also sensitive to the underlying data regime: in high-dimensional settings with limited sample sizes and sparse signals, these methods may exhibit reduced stability and calibration, whereas in settings with stronger nonlinear structure or larger sample sizes, they may offer competitive advantages. These findings underscore that the choice of method should depend on the evaluation criterion most relevant to the analysis, as well as the alignment between model flexibility, sample size, and the underlying data-generating structure.

Note that the comparative results should be interpreted in terms of the specific implementations and tuning strategies employed in this study, rather than as definitive statements about the intrinsic performance of the broader methodological classes. In particular, methods such as DH may exhibit improved performance under more extensive hyperparameter tuning.

From a theoretical perspective, penalized regression methods for survival and competing risks have been extensively studied in high-dimensional settings. Classical results establish the oracle properties of SCAD and MCP \citep{fan_li_2001,zhang_2010} and the consistency of Lasso-type estimators under sparsity assumptions \citep{zhang_lu_2007, huang_et_al_2013}. These results help explain some of our empirical findings—for example, the relative stability of SCAD and MCP in calibration and the limitations of Lasso-like methods in $n<p$ scenarios. While our study was not designed to provide new theoretical insights, connecting the simulation outcomes to existing theory underscores that the observed patterns are consistent with known methodological properties, thereby strengthening confidence in our comparative conclusions.

The observed advantage of CB under the considered settings can be understood from its methodological design: by iteratively updating covariate effects within the Fine–Gray framework, CB stabilizes estimation while maintaining sparsity, thus achieving reliable variable selection and strong discriminative ability in high-dimensional settings. In contrast, RF and DH are highly flexible but less tailored to the competing risks context. RF captures nonlinearities through recursive partitioning, but in sparse high-dimensional data, this often leads to a diluted signal and inflated false discoveries. DH, while capable of modeling complex dependencies, is sensitive to network architecture and hyperparameter choices, which can yield instability and poor calibration under limited sample sizes. These contrasts suggest that CB's balance of model-based structure and iterative refinement makes it more robust across diverse scenarios, whereas RF and DH may require larger data or more specialized tuning to realize their full potential.

Importantly, this design implies that the comparison is not fully neutral across modeling paradigms: methods that directly target the subdistribution hazard (e.g., PR and CB) are structurally aligned with the data-generating mechanism, whereas methods based on alternative formulations (e.g., RF and DH) may be comparatively disadvantaged, even if they are capable of capturing more general nonlinear or nonproportional effects.

Several limitations must be acknowledged due to the scale and complexity of our analysis. As our simulations were based on the Fine--Gray model, the design may be relatively favorable to PR and CB approaches. While the Fine--Gray model represents a widely used and practically relevant setting, it also implies that the comparison is not fully neutral across modeling paradigms: methods that directly target the subdistribution hazard (e.g., PR and CB) are structurally aligned with the data-generating mechanism, whereas methods based on alternative formulations (e.g., RF and DH) may be comparatively disadvantaged, even if they are capable of capturing more general nonlinear or nonproportional effects. Consequently, observed performance differences should be interpreted as reflecting both methodological properties and the degree of alignment with the underlying data-generating framework. Alternative data-generating mechanisms (e.g., cause-specific hazards, nonproportional effects, or more complex dependent censoring structures) may lead to different performance patterns, and the extent to which different methods maintain their relative performance under such departures remains an important open question. The performance of methods is influenced by hyperparameter tuning, and default settings may not fully optimize each approach. Deep learning methods, in particular, are highly sensitive to network architecture, which was not exhaustively explored in this study. Thus, while our findings suggest poor calibration for DH under the implemented settings, results may vary under alternative architectures or more extensive tuning. While we evaluated independent, exchangeable, and AR(1) correlation structures, real-world data may exhibit more intricate dependency patterns that were not fully captured \citep{becker_et_al_2023}. As an illustrative use-case, in the melanoma application, we did not assess clinical utility metrics (e.g., decision curve analysis) or perform resampling-based or external validation, which would provide further insight into real-world applicability \citep{marchetti_et_al_2021}. Additionally, our study assumes complete case analysis or simple imputation, whereas real-world competing risk datasets often contain missing covariates, which can significantly impact model performance \citep{carroll_et_al_2020}. Finally, our comparison focused on non-Bayesian methods, leaving Bayesian approaches for future investigation \citep{monterrubio_et_al_2024}.

An important consideration is that our evaluation focused on the direct application of commonly used high-dimensional methods, without incorporating preliminary variable screening or dimension-reduction strategies. Prior work has shown that screening approaches (e.g., sure independence screening \citep{fan_lv_2008}, conditional screening \citep{hong_et_al_2018}) or projection methods (e.g., principal components, partial least squares) can effectively reduce the covariate space, potentially stabilizing penalized regression and mitigating some of the instabilities we observed. While these strategies may improve performance in practice, they also introduce additional tuning and model-selection complexity. Our comparative framework, therefore, highlights the baseline performance of methods applied ``as is,'' but future research should explore how integrating screening or dimension-reduction with penalized or machine-learning approaches may yield more robust results in challenging $n<p$ settings.

Our findings suggest several promising directions for future research. Extending the present study to incorporate alternative data-generating scenarios mentioned above would provide a more comprehensive assessment of robustness, but would substantially increase the scope and computational burden, given the already extensive simulation design. We therefore view such investigations as a natural and important direction for future work. Efforts should be made to enhance calibration techniques for CB, improve the interpretability of nonlinear methods such as RF and DH, and explore hybrid approaches that balance variable selection accuracy, discrimination, and calibration in high-dimensional competing risk settings. While IBS provides an integrated measure of prediction error over time, complementary graphical tools such as calibration plots at selected time points may offer additional insight into model calibration and are left for future investigation. Additionally, a comparative analysis of Bayesian competing risk methods in high-dimensional data remains an important avenue for future exploration.

%%%%%%%%%%%%%%%%%%%%%%%%%%%%%%%%%%%%%%%%%%%%%%
%% Example with single Appendix:            %%
%%%%%%%%%%%%%%%%%%%%%%%%%%%%%%%%%%%%%%%%%%%%%%
%\begin{appendix}
%\section*{Title}\label{appn} %% if no title is needed, leave empty \section*{}.
%Appendices should be provided in \verb|{appendix}| environment,
%before Acknowledgements.
%
%If there is only one appendix,
%then please refer to it in text as \ldots\ in the \hyperref[appn]{Appendix}.
%\end{appendix}
%%%%%%%%%%%%%%%%%%%%%%%%%%%%%%%%%%%%%%%%%%%%%%
%% Example with multiple Appendixes:        %%
%%%%%%%%%%%%%%%%%%%%%%%%%%%%%%%%%%%%%%%%%%%%%%
%\begin{appendix}
%\section{Title of the First Appendix}\label{appA}
%If there are more than one appendix, then please refer to it
%as \ldots\ in Appendix \ref{appA}, Appendix \ref{appB}, etc.
%
%\section{Title of the Second Appendix}\label{appB}
%\subsection{First Subsection of Appendix \protect\ref{appB}}
%
%Use the standard \LaTeX\ commands for headings in \verb|{appendix}|.
%Headings and other objects will be numbered automatically.
%$$
%\mathcal{P}=(j_{k,1},j_{k,2},\dots,j_{k,m(k)}). \label{path}
%$$
%
%Sample of cross-reference to the formula (\ref{path}) in Appendix \ref{appB}.
%\end{appendix}

%%%%%%%%%%%%%%%%%%%%%%%%%%%%%%%%%%%%%%%%%%%%%%
%% Support information, if any,             %%
%% should be provided in the                %%
%% Acknowledgements section.                %%
%%%%%%%%%%%%%%%%%%%%%%%%%%%%%%%%%%%%%%%%%%%%%%
\section*{Author Contributions}
All authors contributed equally to this work.

\section*{Acknowledgments}
The authors thank the JAKAR High-Performance Cluster at the University of Texas at El Paso for providing computational resources free of charge.

\section*{Supplemenary Material} 
\textbf{Supplement to ``Comparative Review of Modern Competing Risk Methods for High-dimensional Data"}

%%%%%%%%%%%%%%%%%%%%%%%%%%%%%%%%%%%%%%%%%%%%%%
%% Funding information, if any,             %%
%% should be provided in the                %%
%% funding section.                         %%
%%%%%%%%%%%%%%%%%%%%%%%%%%%%%%%%%%%%%%%%%%%%%%
%\begin{funding}
%The first author was supported by NSF Grant DMS-??-??????.
%
%The second author was supported in part by NIH Grant ???????????.
%\end{funding}

%%%%%%%%%%%%%%%%%%%%%%%%%%%%%%%%%%%%%%%%%%%%%%
%% Supplementary Material, including data   %%
%% sets and code, should be provided in     %%
%% {supplement} environment with title      %%
%% and short description. It cannot be      %%
%% available exclusively as external link.  %%
%% All Supplementary Material must be       %%
%% available to the reader on Project       %%
%% Euclid with the published article.       %%
%%%%%%%%%%%%%%%%%%%%%%%%%%%%%%%%%%%%%%%%%%%%%%

\section*{Supplemenary Material} 
\textbf{Supplement to "Comparative Review of Modern Competing Risk Methods for High-dimensional Data" by Paul M. Djangang, Summer S. Han, and Nilotpal Sanyal}

%\begin{supplement}
%\stitle{Title of Supplement B}
%\sdescription{Short description of Supplement B.}
%\end{supplement}

\bibliographystyle{jasa}
\bibliography{mybib}
\end{document}